\documentclass[preprint,showpacs,preprintnumbers,amsmath,amssymb,nofootinbib]{revtex4}
\usepackage{graphicx,subfigure,textcomp}
\usepackage{dcolumn}
\usepackage{bm}
\usepackage{mathrsfs,slashed,amsmath}
\usepackage{natbib}
\allowdisplaybreaks
\begin{document}
\title{Chiral extrapolation of baryon mass ratios}
\author{Peter C. Bruns}
\affiliation{Institut f\"ur Theoretische Physik, Universit\"at Regensburg, D-93040 Regensburg, Germany}
\author{Ludwig Greil}
\affiliation{Institut f\"ur Theoretische Physik, Universit\"at Regensburg, D-93040 Regensburg, Germany}
\author{Andreas Sch\"afer}
\affiliation{Institut f\"ur Theoretische Physik, Universit\"at Regensburg, D-93040 Regensburg, Germany}
\date{\today}
\begin{abstract}
We analyze lattice data for octet baryon masses from the QCDSF collaboration employing manifestly covariant Baryon Chiral Perturbation Theory. It is shown that certain combinations of low-energy constants can be fixed more accurately than before from this data. We also examine the impact of this analysis on the pion-nucleon sigma term, and on the convergence properties of baryon mass expansions in the SU(3) symmetry limit.
\end{abstract}
\maketitle
\section{Introduction}
\label{sec:Intro}

Three-flavor Baryon Chiral Perturbation Theory (BChPT) \cite{Jenkins:1990jv,Krause:1990xc} is an effective field theory for QCD in the low-energy regime where the lowest-lying meson-octet and the lowest-lying baryon octet (and possibly also the decuplet) are the relevant degrees of freedom. 

It is now known for many years that chiral expansions derived from this framework are usually not under sufficient theoretical control because of an unsatisfying convergence behavior\footnote{Of course, we are not concerned here with convergence in the strict mathematical sense, but with the reliability of theoretical predictions at low orders in perturbation theory.}: higher-order corrections from chiral loops can in general not be guaranteed to be much smaller than the leading terms. The quark mass expansion around the three-flavor chiral limit converges slowly because the strange quark mass, $m_{s}$, is much larger than the light quark masses $m_{\ell}=m_{u,d}$, and induces sizeable kaon loop corrections to the leading tree level results. Detailed discussions on the convergence behavior of BChPT from various perspectives can be found in \cite{Borasoy:1996bx,Gasser:1997rx,Donoghue:1998rp,Donoghue:1998bs,Mojzis:1999qw,Djukanovic:2006xc,Mai:2009ce,Meissner:1994wy,WalkerLoud:2008bp,Ishikawa:2009vc}. The essential point, however, can be seen immediately from a comparison of mass scales: while $\sqrt{2B_{0}m_{\ell}}\sim M_{\pi}\approx 140\,\mathrm{MeV}$ is small compared to a typical hadronic scale of $\sim 1\,\mathrm{GeV}$, the three-flavor expansions can involve $\sqrt{2B_{0}m_{s}}=: M_{\bar{s}s}\approx 700\,\mathrm{MeV}$, which is not small compared to that reference scale. In fact, loop graphs involving the eta meson are only suppressed by an additional factor of $M_{\eta}/(4\pi F_{0})\approx 0.5$, which can easily be compensated by additional prefactors ($F_{0}$ is the meson decay constant in the three-flavor chiral limit, while $B_{0}$ is related to the quark condensate in the same limit). 

The problem is particularly urgent for chiral extrapolation formulae for lattice simulations. There, the meson masses are in general even larger than for physical quark masses. Consequently, the application of BChPT to such lattice data is not under satisfactory theoretical control. Though one can obtain fits that seem to describe the data fairly well, a stable determination of the corresponding low-energy constants (LECs) is not yet within reach, because the extracted values can be strongly affected by uncontrolled higher-order corrections. For some specific application, one might not even worry much about this, but one should recall that the main point of ChPT is to yield relations between many different observables, in terms of a limited set (at least at a fixed order in the low-energy expansion) of parameters. For example, the baryon mass in the chiral limit, $m_{0}$, and its leading quark mass corrections will unavoidably appear in practically {\em any} complete loop calculation in BChPT, in baryon form factors, meson-baryon scattering lengths, etc. Therefore, one should try to somehow control the higher-order loop corrections.

Many alternative ways to overcome this problem have been proposed in the literature, e.g. long-distance regularization \cite{Donoghue:1998bs}, reordering and resummation schemes \cite{Mojzis:1999qw,Semke:2005sn}, large-$N_{c}$-scaling arguments \cite{FloresMendieta:2000mz,WalkerLoud:2011ab,Cordon:Aug2012,Lutz:2012mq} and two-flavor expansions for hyperon properties \cite{Tiburzi:2008bk,Jiang:2009fa}, each having its own merits. In this work, we stick to a standard scheme (evaluating the chiral loops employing infrared regularization \cite{Becher:1999he}) and thus analyze the quark mass expansions of the baryon octet masses in a fashion closely following \cite{Frink:2004ic}. We refer to that reference for most of the corresponding technical details (see also \cite{Ellis:1999jt}). For studies in different schemes, see e.g. \cite{Borasoy:1996bx,Jenkins:1991ts,Bernard:1993nj,Lebed:1994gt,WalkerLoud:2004hf,Frink:2005ru,Lehnhart:2004vi,Young:2009zb,MartinCamalich:2010fp,Durr:2011mp,Semke:2011ez,Semke:2012gs,Ren:2012aj}. We have reproduced the complete one-loop ($\mathcal{O}(p^{4})$) calculation in this covariant scheme. Our aim here is to put to use calculations like \cite{Frink:2004ic}, by analyzing the lattice data presented in \cite{Bietenholz:2011qq} which covers a quark mass region where all the masses of the pseudoscalar bosons are {\em smaller} than the physical eta mass, and to give improved estimates for LECs featuring in these calculations. We consider this a first and necessary step toward a controlled application of BChPT to such data sets. While in most three-flavor lattice simulations, the (large) strange quark mass is fixed\footnote{In \cite{Aoki:2008sm}, the strange quark mass was $20\%$ larger than the physical value, resulting in large kaon and eta masses, which makes the application of three-flavor chiral extrapolations even more problematic.}, and the physical point is approached lowering $m_{\ell}$ (see e.g. \cite{Aoki:2008sm,Durr:2010aw} for recent examples), the strategy in \cite{Bietenholz:2011qq} is different: One simulates at a ``symmetric point'' where $m_{\ell}=m_{s}$, and then approaches the physical point increasing the quark mass difference, but keeping the average quark mass fixed. At the symmetric point, the pseudoscalar octet mesons all have the same mass of about $400\,\mathrm{MeV}$, and on the trajectory to the physical point, the kaons and etas are lighter than observed in nature (there even exist some data points where the meson masses are all $< 400\,\mathrm{MeV}$ at $m_{s}=m_{\ell}$). This is already a nice feature, but there is a second point that is probably even more important. In a surprisingly large region around the symmetric point, approximations neglecting all but linear symmetry-breaking effects were shown in \cite{Bietenholz:2011qq} to yield a very good description of the data. In particular, quantities which do not receive linear symmetry-breaking contributions are observed to be very stable even down to the physical point (one example for such a quantity is the famous Gell-Mann-Okubo mass difference). This behavior can not be understood from ChPT alone, but it puts tight constraints on certain combinations of LECs (e.g. by enforcing large cancellations in higher-order terms in the $\delta m_{\ell}$-expansion of such quantities). It is convenient for the purpose of studying the symmetry-breaking effects separately to analyze certain baryon mass {\em ratios}, where a large factor with the dimension of mass (essentially the baryon mass at the symmetric point) cancels out. The behavior of these ratios can be very well described by BChPT (see our Fig.~\ref{fig:fanplot3sets}, and also \cite{Shanahan:2012wh} for a corresponding illustration). Here we want to exploit the advantages just mentioned for the determination of the LEC-combinations occurring in octet baryon masses, in the hope that this will help e.g. when analyzing lattice data for other quantities with the same strategy.

This paper is organized as follows: After briefly introducing some basic notation in sec.~\ref{sec:Vorspiel}, we collect the input from the purely mesonic sector in sec.~\ref{sec:Mesons} in some detail. This is necessary, because the quark-mass dependence of the masses of the pseudoscalar mesons is needed to rewrite the quark mass expansions of the baryon masses at $\mathcal{O}(p^{4})$ in terms of the meson masses. We also include some simple examples of our strategy of analysis in this section. In sec.~\ref{sec:q3}, we consider the result of the leading one-loop ($\mathcal{O}(p^{3})$) calculation, to set the stage for the later developments. We also show some examples of a first numerical analysis, analyzing the experimental baryon masses. Sec.~\ref{sec:q4} completes the outline of our framework specifying the Lagrangians and diagrams entering at fourth chiral order. Readers who are only interested in the results for the baryon sector are invited to skip secs.~\ref{sec:Mesons}-\ref{sec:q4} and directly continue with sec.~\ref{sec:Analyse}, in which we discuss in detail the extrapolation functions we use, and which combinations of LECs can be extracted more accurately than before from the data we consider (and which can not). In a series of tables, we present our fit results, obtained with different fit strategies and input parameter sets. We discuss our results in sec.~\ref{sec:Diskussion} and give a short conclusion of our findings.

\section{Preliminaries}
\label{sec:Vorspiel}
Throughout we work in the isospin limit and set the $\ell$ight quark masses to $m_{u}=m_{d}=:m_{\ell}$. In \cite{Bietenholz:2011qq}, the quark masses are varied along a trajectory $T$ in the $(m_{\ell},m_{s})$-space where the average quark mass $\bar{m}:=\frac{1}{3}(2m_{\ell}+m_{s})$ is kept fixed, while $\delta m_{\ell}=m_{\ell}-\bar{m}\,$ is varied. The aforesaid trajectory $T$  connects the symmetric point, where $m_{\ell}=m_{s}=\bar{m}$, and the physical point where the strange quark mass attains its physical value while $m_{\ell}=\frac{1}{2}(m_{u}+m_{d})$.\\ To get familiar with the notation, let us consider the leading terms in the quark-mass expansion of the squares of the pseudo-Goldstone-boson masses. We have
\begin{align}
\dot{M}_{\pi}^{2} &:= 2B_{0}m_{\ell} = 2B_{0}\bar{m}+2B_{0}\delta m_{\ell}\,,\\
\dot{M}_{K}^{2} &:= B_{0}(m_{\ell}+m_{s}) = 2B_{0}\bar{m}-B_{0}\delta m_{\ell}\,,\\
\dot{M}_{\eta}^{2} &:= \frac{2}{3}B_{0}(m_{\ell}+2m_{s}) = 2B_{0}\bar{m}-2B_{0}\delta m_{\ell}\,.
\end{align}
Here $B_{0}$ is related to the quark condensate in the chiral limit, see e.g. \cite{Gasser:1984gg}. The leading term of the pseudo-Goldstone-boson mass at the symmetric point is then found by setting $B_{0}\delta m_{\ell}=0$,
\begin{align}\label{eq:MDotstar}
\dot{M}_{\star}^{2} = 2B_{0}\bar{m}\,.
\end{align}
Our notation is set up such that quantities in the chiral limit will be denoted by a subscript $0$, while quantities at the symmetric point on $T$ are marked with a $\star$\,.\\In \cite{Bietenholz:2011qq}, it was noted that some specific combinations of hadron masses are very stable when varying the quark masses along $T$, compare e.g. Tab.~2 and Fig.~13 in that paper. Below, we list three of these combinations which are relevant for our present study:
\begin{align}
X_{N} &= \frac{1}{3}(m_{N}+m_{\Sigma}+m_{\Xi}) \approx 1150\,\mathrm{MeV}\,,\label{eq:XN}\\
X_{\pi}^{2} &= \frac{1}{3}(2M_{K}^{2}+M_{\pi}^{2})\approx (412\,\mathrm{MeV})^{2}\,,\label{eq:Xpi}\\
X_{\eta}^{2} &= \frac{1}{2}(M_{\pi}^{2}+M_{\eta}^{2})\approx(400\,\mathrm{MeV})^{2}\,.\label{eq:Xeta}
\end{align}
The experimental values are given in round brackets.
\section{Meson masses and decay constants to one loop}
\label{sec:Mesons}
\subsection*{Meson masses}
The one-loop expressions for the masses of the pions, kaons and the eta can be found e.g. in \cite{Gasser:1984gg}. We give here the expansions of these formulae in terms of $\bar{m}$ and $\delta m_{\ell}$.
The meson mass squared at the symmetric point is given (to one-loop accuracy) by
\begin{align}\label{eq:Mstar}
M_{\star}^{2} = 2B_{0}\bar{m}\left(1 + \frac{2B_{0}\bar{m}}{(4\pi F_{0})^{2}}\left(128\pi^{2}(-3L_{4}-L_{5}+6L_{6}+2L_{8})+\frac{2}{3}\log\left(\frac{\sqrt{2B_{0}\bar{m}}}{\mu}\right)\right)\right)\,.
\end{align}
The expansion of the meson masses in the symmetry-breaking parameter $\delta m_{\ell}$ reads
\begin{align}
\begin{split}
M_{\pi}^{2} &= M_{\star}^{2}+(2B_{0}\delta m_{\ell})\left(1+\frac{2B_{0}\bar{m}}{(4\pi F_{0})^{2}}\left(\frac{2}{3}+2\log\left(\frac{\sqrt{2B_{0}\bar{m}}}{\mu}\right)-128\pi^{2}\left(3L_{4}+2L_{5}-6L_{6}-4L_{8}\right)\right)\right)\\
            &\quad+ (B_{0}\delta m_{\ell})^{2}\left(\frac{5+8\log\left(\frac{\sqrt{2B_{0}\bar{m}}}{\mu}\right)-768\pi^{2}(L_{5}-2L_{8})}{24\pi^{2}F_{0}^{2}}\right)\,,
\end{split}\\
\begin{split}
M_{K}^{2} &= M_{\star}^{2}+(2B_{0}\delta m_{\ell})\left(-\frac{1}{2}-\frac{2B_{0}\bar{m}}{(4\pi F_{0})^{2}}\left(\frac{1}{3}+\log\left(\frac{\sqrt{2B_{0}\bar{m}}}{\mu}\right)-64\pi^{2}\left(3L_{4}+2L_{5}-6L_{6}-4L_{8}\right)\right)\right) \\ 
 &\quad+ (B_{0}\delta m_{\ell})^{2}\left(\frac{1+\log\left(\frac{\sqrt{2B_{0}\bar{m}}}{\mu}\right)-96\pi^{2}(L_{5}-2L_{8})}{12\pi^{2}F_{0}^{2}}\right)\,,
\end{split}\\
\begin{split}
M_{\eta}^{2} &= M_{\star}^{2}-(2B_{0}\delta m_{\ell})\left(1+\frac{2B_{0}\bar{m}}{(4\pi F_{0})^{2}}\left(\frac{2}{3}+2\log\left(\frac{\sqrt{2B_{0}\bar{m}}}{\mu}\right) - 128\pi^{2}\left(3L_{4}+2L_{5}-6L_{6}-4L_{8}\right)\right)\right) \\
 &\quad- (B_{0}\delta m_{\ell})^{2}\left(\frac{8+12\log\left(\frac{\sqrt{2B_{0}\bar{m}}}{\mu}\right)+768\pi^{2}(L_{5}-12L_{7}-6L_{8})}{24\pi^{2}F_{0}^{2}}\right)\,.
\end{split}
\end{align}
Consequently, we have for the combinations $X_{\pi}^{2},\,X_{\eta}^{2}$:
\begin{align}
X_{\pi}^{2} &= \frac{1}{3}(2M_{K}^{2}+M_{\pi}^{2})= M_{\star}^{2}+\frac{(B_{0}\delta m_{\ell})^{2}}{24\pi^{2}F_{0}^{2}}\left(3+4\log\left(\frac{\sqrt{2B_{0}\bar{m}}}{\mu}\right)-384\pi^{2}(L_{5}-2L_{8})\right)\,,\label{eq:XpiSer}\\
\begin{split}
X_{\eta}^{2} &= \frac{1}{2}(M_{\pi}^{2}+M_{\eta}^{2})\\ 
             &= M_{\star}^{2}-\frac{(B_{0}\delta m_{\ell})^{2}}{48\pi^{2}F_{0}^{2}}\left(3+4\log\left(\frac{\sqrt{2B_{0}\bar{m}}}{\mu}\right)+1536\pi^{2}(L_{5}-6L_{7}-4L_{8})\right)\,.\label{eq:XetaSer}
\end{split}
\end{align}
The low-energy constants $L_{i}$ of course correspond to the renormalized constants $L_{i}^{r}(\mu)$ of \cite{Gasser:1984gg}. In \cite{Bietenholz:2011qq}, instead of the quark mass difference $B_{0}\delta m_{\ell}$, the variable
\begin{align}
\begin{split}
\nu &= \frac{M_{\pi}^{2}-X_{\pi}^{2}}{X_{\pi}^{2}} \label{eq:nu} \\ 
    &=\frac{2B_{0}\delta m_{\ell}}{M_{\star}^{2}}\left(1+\frac{2B_{0}\bar{m}}{24\pi^{2}F_{0}^{2}}\left(1+3\log\left(\frac{\sqrt{2B_{0}\bar{m}}}{\mu}\right)-192\pi^{2}(3L_{4}+2L_{5}-6L_{6}-4L_{8})\right)\right)\\ 
    &\quad+\frac{(2B_{0}\delta m_{\ell})^{2}}{48\pi^{2}F_{0}^{2}M_{\star}^{2}}\left(1+2\log\left(\frac{\sqrt{2B_{0}\bar{m}}}{\mu}\right)-192\pi^{2}(L_{5}-2L_{8})\right)\\
    &\quad+\mathcal{O}(\bar{m}^{2}\delta m_{\ell},\bar{m}(\delta m_{\ell})^{2},(\delta m_{\ell})^{3})
\end{split}
\end{align}
is used to parameterize the SU(3) symmetry breaking. Inverting this relation and again expanding up to the second order in the symmetry-breaking variable $\delta m_{\ell}$, we get
\begin{align}
\begin{split}
2B_{0}\delta m_{\ell} &= \nu M_{\star}^{2}\left(1-\frac{2B_{0}\bar{m}}{24\pi^{2}F_{0}^{2}}\left(1+3\log\left(\frac{\sqrt{2B_{0}\bar{m}}}{\mu}\right)-192\pi^{2}(3L_{4}+2L_{5}-6L_{6}-4L_{8})\right)\right)\\
                      &\quad- \frac{(\nu M_{\star}^{2})^{2}}{48\pi^{2}F_{0}^{2}}\left(1+2\log\left(\frac{\sqrt{2B_{0}\bar{m}}}{\mu}\right)-192\pi^{2}(L_{5}-2L_{8})\right)\,,\label{eq:nuRel}
\end{split}
\end{align}
where we have also truncated the chiral expansion of the coefficient functions in this series.
\subsection*{Decay constants}
Similarly, the expansion of the decay constants is given by
\begin{align}
F_{\pi} &= F_{\star}-(B_{0}\delta m_{\ell})\left(\frac{3+6\log\left(\frac{\sqrt{2B_{0}\bar{m}}}{\mu}\right)-256\pi^{2}L_{5}}{32\pi^{2}F_{0}}\right) +\mathcal{O}((\delta m_{\ell})^{2})\,,\\
F_{K} &= F_{\star}+(B_{0}\delta m_{\ell})\left(\frac{3+6\log\left(\frac{\sqrt{2B_{0}\bar{m}}}{\mu}\right)-256\pi^{2}L_{5}}{64\pi^{2}F_{0}}\right) +\mathcal{O}((\delta m_{\ell})^{2})\,,\\
F_{\eta} &= F_{\star}+(B_{0}\delta m_{\ell})\left(\frac{3+6\log\left(\frac{\sqrt{2B_{0}\bar{m}}}{\mu}\right)-256\pi^{2}L_{5}}{32\pi^{2}F_{0}}\right) +\mathcal{O}((\delta m_{\ell})^{2})\,,
\end{align}
where
\begin{align}\label{eq:Fstar}
F_{\star} = F_{0}\left(1+\frac{2B_{0}\bar{m}}{(4\pi F_{0})^{2}}\left(64\pi^{2}(3L_{4}+L_{5})-3\log\left(\frac{\sqrt{2B_{0}\bar{m}}}{\mu}\right)\right)\right)+\mathcal{O}(\bar{m}^{2})\,.
\end{align}
Note that if terms nonlinear in the symmetry-breaking variable are neglected, the combination 
\begin{align}\label{eq:FX}
F_{X}:=\frac{1}{3}(F_{\pi}+2F_{K})
\end{align}
is stable at that order.
\subsection*{Quark masses}
For the sake of completeness, we also give an expression for the quantity $M_{\bar{s}s}^{2}:=2B_{0}m_{s}$ in terms of the one-loop formulae for the meson masses:
\begin{align}
\begin{split}
M_{\bar{s}s}^{2} &= 2\dot{M}_{K}^{2}-\dot{M}_{\pi}^{2} = 2M_{K}^{2}-M_{\pi}^{2}\\ 
                 &\quad+ \frac{1}{48\pi^{2}F_{0}^{2}}\biggl(384\pi^{2}\left(M_{K}^{4}(4L_{4}+2L_{5}-8L_{6}-4L_{8}) - M_{\pi}^{4}(L_{4}+L_{5}-2L_{6}-2L_{8})\right)\\ 
                 &\quad+ 3M_{\pi}^{4}\log\left(\frac{M_{\pi}}{\mu}\right)-(3M_{\eta}^{4}+2M_{\eta}^{2}M_{\pi}^{2})\log\left(\frac{M_{\eta}}{\mu}\right)\biggr)+\mathcal{O}(M^{6})\,.\label{eq:MSS} 
\end{split}
\end{align}
Keeping this quantity fixed, one can express the light quark mass dependence of the baryon masses through the pion mass dependence, using
\begin{align}
\begin{split}
2B_{0}m_{\ell} &= M_{\pi}^{2}\biggl(1+\frac{1}{144\pi^{2}F_{0}^{2}}\biggl(1152\pi^{2}\left(M_{\pi}^{2}(2L_{4}+L_{5}-4L_{6}-2L_{8})+M_{\bar{s}s}^{2}(L_{4}-2L_{6})\right)\\ 
               &\quad+ (M_{\pi}^{2}+2M_{\bar{s}s}^{2})\log\left(\frac{\sqrt{M_{\pi}^{2}+2M_{\bar{s}s}^{2}}}{\sqrt{3}\mu}\right)-9M_{\pi}^{2}\log\left(\frac{M_{\pi}}{\mu}\right)\biggr)\biggr)+\mathcal{O}(M^{6})\,.\label{eq:2Bmell} 
\end{split}
\end{align}
\subsection*{Numerical analysis}
For an overview on lattice determinations of the parameters $F_{0},L_{i}$ we refer to the recent discussion in \cite{Colangelo:2010et}. We reproduce three sets of results in the table below. As central values in our numerical analysis, we adopt the set of parameters displayed in the first row (MILC2010). The other three parameter sets are used only for the error estimates of our results. The numerical values of the $L_{i}$ are given at a renormalization scale of $\mu=770\,\mathrm{MeV}$.
\begin{table}[h]
\caption{Three parameter sets for the low-energy coefficients $F_{0},L_{i}$.}
\label{tab:Li}
\begin{center}
\begin{ruledtabular}
\begin{tabular}{c c c c c c}
$F_{0}$ (MeV) & $10^{3}L_{4}$ & $10^{3}L_{5}$ & $10^{3}L_{6}$ & $10^{3}L_{8}$ & Ref. \\
\hline
$80.3\pm 6.0$ & $-0.08\pm 0.60$ & $0.98\pm 0.40$ & $-0.02\pm 0.40$ & $0.42\pm 0.30$ & MILC2010 \cite{Bazavov:2010hj} \\ 
\hline
$78.3\pm 3.2$ & $0.04\pm 0.14$ & $0.84\pm 0.38$ & $0.07\pm 0.11$ & $0.36\pm 0.09$ & MILC2009 \cite{Bazavov:2009fk} \\
\hline
$83.8\pm 6.4$ & $-0.06\pm 0.10$ & $1.45\pm 0.07$ & $0.02\pm 0.05$ & $0.62\pm 0.04$ & PACS2008 \cite{Aoki:2008sm} \\
\end{tabular} 
\label{tab:LECS}
\end{ruledtabular}
\end{center}
\end{table}
At the order we are working, we can replace $F_{0}$ in eq.~(\ref{eq:Mstar}) by $F_{\star}$, see eq.~(\ref{eq:Fstar}). Given the observed stability of $X_{\pi}$, and the fact that this quantity is even used to set the scale for meson observables in \cite{Bietenholz:2011qq}, we determine the parameter $2B_{0}\bar{m}$ from the requirement that $M_{\star}\overset{!}{=}X_{\pi}^{phys}=412\,\mathrm{MeV}$. We find
\begin{align}
\sqrt{2B_{0}\bar{m}} \equiv \dot{M}_{\star} &\overset{!}{=} (420\pm 40)\,\mathrm{MeV}\,,\label{eq:Mstarnum}\\
\Rightarrow\,F_{\star} &= (112\pm 15)\,\mathrm{MeV}\,.\label{eq:Fstarnum}
\end{align}
We note that inserting the value for $F_{0}$ directly in eq.~(\ref{eq:Mstar}), one finds very similar values, $\dot{M}_{\star} = 428\,\mathrm{MeV}$ and $F_{\star}=112.5\,\mathrm{MeV}$, while the experimental value for $F_{X}$ is $F_{X}^{phys}=105\,\mathrm{MeV}$. As the one-loop effects discussed above enter our calculation of baryon masses formally at NNLO (meson masses) or even at two-loop order (decay constants), the estimates given in  eqs.~(\ref{eq:Mstarnum}),(\ref{eq:Fstarnum}) are sufficient for our purposes. \\
We also note that the fan plots for the meson masses shown in \cite{Bietenholz:2011qq} can be nicely reproduced with the values of the LECs employed above, while $X_{\pi}$ is indeed almost a constant along $T$. For this to be the case, one should expect from eq.~(\ref{eq:XpiSer}) that 
\begin{align}
384\pi^{2}(L_{5}-2L_{8}) \approx 3+4\log\left(\frac{420}{770}\right) \sim 0.58.
\end{align}
In fact, for our central values (MILC2010) we have 
\begin{align}
384\pi^{2}(L_{5}(\mu=770\,\mathrm{MeV})-2L_{8}(\mu=770\,\mathrm{MeV}))\sim 0.53\,,
\end{align}
so that terms in $X_{\pi}$ quadratic in the symmetry breaking are indeed tiny. For $X_{\eta}$ to be equally stable, we infer from eq.~(\ref{eq:XetaSer}) that we should have $L_{7}\sim -0.1\cdot 10^{-3}$.
\begin{figure}[h]
\centering
\includegraphics[width=10.5cm]{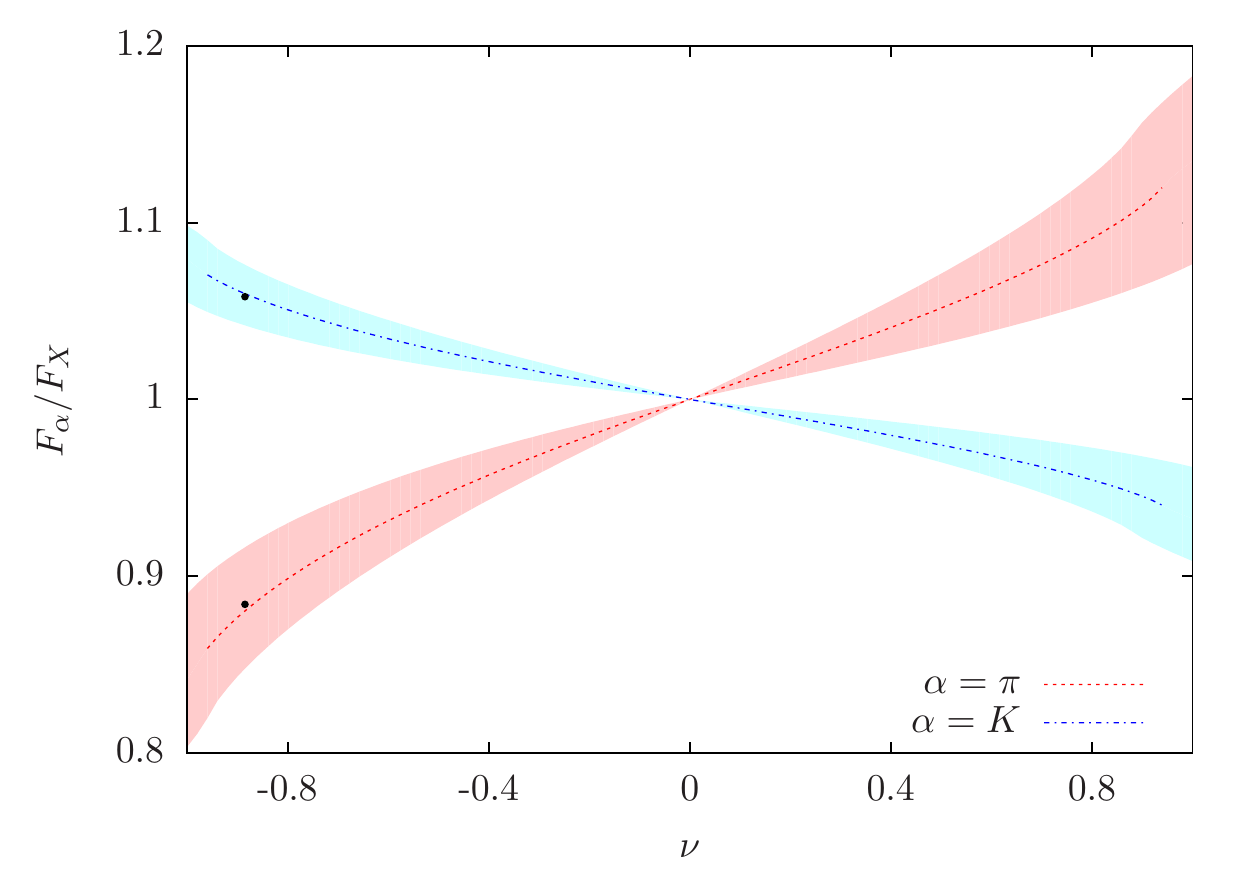} 
\caption{A ``fan plot'' for the symmetry-breaking effects in the meson decay constants $F_{K}$ and $F_{\pi}$. The black dots denote the experimental values.}
\label{fig:ffanplot}
\end{figure}
In Fig.~\ref{fig:ffanplot}, we show a ``fan plot'' for the meson decay constants, i.e. we plot the ratios $F_{\pi}/F_{X}$ and $F_{K}/F_{X}$ as functions of the variable $\nu$ which parametrizes the SU(3) symmetry breaking, see eq.~(\ref{eq:nuRel}). The derivative at the symmetric point ($\nu=0$) is essentially determined by the LEC $L_{5}$, which we vary within the error according to MILC2010 to generate the bands around the central curves. One observes that the fan plot shows a linear behavior of the ratios with $\nu$ in the vicinity of the symmetric point, but some curvature is visible near the physical point, which seems to be necessary to reproduce the experimental values of $F_{\pi}$ and $F_{K}$.\\ Inserting the experimental pion and kaon masses in eq.~(\ref{eq:MSS}), one arrives at
\begin{align}\label{eq:MSSnum}
M_{\bar{s}s}=(715\pm 100)\,\mathrm{MeV}\,,
\end{align}
where again we use the MILC2010 parameters as central values. The uncertainty due to the variation of the input parameters is somewhat larger than in eq.~(\ref{eq:Mstarnum}). We have a clear hierarchy
\begin{align}
M_{\pi}\approx \sqrt{2B_{0}m_{\ell}}<\sqrt{2B_{0}\bar{m}}<\sqrt{2B_{0}m_{s}}=M_{\bar{s}s}\,,
\end{align}
compare eqs.~(\ref{eq:Mstarnum}) and (\ref{eq:MSSnum}). While it is quite obvious that an expansion in $M_{\bar{s}s}$ (over a typical hadronic scale) will be ineffective, an expansion in $\dot{M}_{\star}$ can be expected to work much better, since the expansion parameter is about $40\%$ smaller.
\section{Baryon masses at third chiral order}
\label{sec:q3}
\begin{figure}[h]
\centering
\includegraphics[width=5cm]{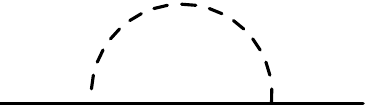} 
\caption{Loop graph with axial meson-baryon vertices from the leading-order chiral Lagrangian, which yields the leading nonanalytic quark mass correction $\sim\sqrt{m_{q}}^{3}$. The full lines represent the baryons, while the dashed line represents the pseudoscalar meson $\pi,K$ or $\eta$.}
\label{fig:q3loop}
\end{figure}

From the relevant terms in the chiral Lagrangian,
\begin{align}
\begin{split}
\mathcal{L}_{\phi B} &= \langle\bar{B}(i\slashed{D}-m_{0})B\rangle +\frac{D}{2}\langle\bar{B}\gamma^{\mu}\gamma_{5}\lbrace u_{\mu},B\rbrace\rangle + \frac{F}{2}\langle\bar{B}\gamma^{\mu}\gamma_{5}\lbrack u_{\mu},B\rbrack\rangle\\ 
&\quad+  b_0 \langle\bar{B} B\rangle \langle\chi_+\rangle + b_{D} \langle\bar{B}\lbrace\chi_+,B\rbrace\rangle + b_{F} \langle\bar{B} \lbrack\chi_+,B\rbrack\rangle + \ldots \label{eq:Lagr12}
\end{split}
\end{align}
one derives the octet baryon masses to  $\mathcal{O}(p^3)$ in BChPT (see e.g. \cite{Bernard:1993nj})
\begin{align}
\begin{split}
m_{N} &= m_{0}-4 b_{D} \dot{M}_{K}^2+4 b_{F} \left(\dot{M}_{K}^2-\dot{M}_{\pi}^2\right)-2 b_{0} \left(2 \dot{M}_{K}^2+\dot{M}_{\pi}^2\right)\\ 
&\quad-\frac{-6D F \left(\dot{M}_{\eta}^3+2 \dot{M}_{K}^3-3 \dot{M}_{\pi}^3\right)+9 F^2 \left(\dot{M}_{\eta}^3+2 \dot{M}_{K}^3+\dot{M}_{\pi}^3\right)+D^2\left(\dot{M}_{\eta}^3+10 \dot{M}_{K}^3+9 \dot{M}_{\pi}^3\right)}{96\pi F_{0}^2 }\,, 
\end{split}\\
\begin{split}
m_{\Lambda} &= m_{0}+\frac{4}{3} b_{D} \left(-4 \dot{M}_{K}^2+\dot{M}_{\pi}^2\right)-2 b_{0} \left(2 \dot{M}_{K}^2+\dot{M}_{\pi}^2\right)-\frac{9F^2\dot{M}_{K}^3+D^2\left(\dot{M}_{\eta}^3+\dot{M}_{K}^3+3\dot{M}_{\pi}^3\right)}{24\pi F_{0}^2}\,, 
\end{split}\\
\begin{split}
m_{\Sigma} &= m_{0}-4 b_{D} \dot{M}_{\pi}^2-2 b_{0} \left(2 \dot{M}_{K}^2+\dot{M}_{\pi}^2\right)-\frac{D^2 \left(\dot{M}_{\eta}^3+3 \dot{M}_{K}^3+\dot{M}_{\pi}^3\right)+3F^2 \left(\dot{M}_{K}^3+2 \dot{M}_{\pi}^3\right)}{24\pi F_{0}^2 }\,, 
\end{split}\\
\begin{split}
m_{\Xi} &= m_{0}-4 b_{D} \dot{M}_{K}^2-4 b_{F} \left(\dot{M}_{K}^2-\dot{M}_{\pi}^2\right)-2 b_{0} \left(2 \dot{M}_{K}^2+\dot{M}_{\pi}^2\right) \\ 
        &\quad-\frac{6 D F \left(\dot{M}_{\eta}^3+2 \dot{M}_{K}^3-3 \dot{M}_{\pi}^3\right)+9 F^2 \left(\dot{M}_{\eta}^3+2 \dot{M}_{K}^3+\dot{M}_{\pi}^3\right)+D^2 \left(\dot{M}_{\eta}^3+10 \dot{M}_{K}^3+9 \dot{M}_{\pi}^3\right)}{96\pi F_{0}^2 }\,, 
\end{split}
\end{align}
from the second order quark mass insertions and the sunrise-type loop graph of Fig.~\ref{fig:q3loop}. To this order, the mass of the eta can be eliminated by using the Gell-Mann-Okubo relation,
\begin{align}
3\dot{M}_{\eta}^{2} = 4 \dot{M}_{K}^{2}-\dot{M}_{\pi}^{2}\,.\label{eq:MesGMO}
\end{align}
Inserting $\dot{M}_{K}\rightarrow 495\,\mathrm{MeV}$ and $\dot{M}_{\pi}\rightarrow 140\,\mathrm{MeV}$ gives $\dot{M}_{\eta}\simeq 566\,\mathrm{MeV}$.
In our numerical work, we shall use $D=\frac{3}{4},\,F=\frac{1}{2}$ which is consistent with the available literature \cite{Jaffe:1989jz,Close:1993mv,Borasoy:1998pe,Lin:2007ap}, see also \cite{WalkerLoud:2011ab}. It is difficult to reliably obtain these two axial coupling constants from a fit to baryon masses. In \cite{WalkerLoud:2011ab}, values close to the expectation from phenomenology were obtained imposing a large-$N_{c}$ counting and restricting the fit to certain mass combinations where the singlet piece drops out. See also \cite{Semke:2012gs} for a similar result. For the above choice of $D,F$, the $p^{3}$-terms cancel completely in the combination
\begin{align}\label{eq:DeltaDF}
\Delta_{DF} := \frac{1}{4}\left(12m_{\Sigma}-4m_{\Lambda}-5m_{N}-3m_{\Xi}\right) = \frac{2}{3}\left(20b_{D}-3b_{F}\right)\left(\dot{M}_{K}^{2}-\dot{M}_{\pi}^{2}\right)+\mathcal{O}(p^{4})\,.
\end{align}
This quantity vanishes in the SU(3) limit, however, it is not too small in the real world. Inserting the physical values for the baryon masses (averaging over isospin multiplets),
\begin{align}\label{eq:mBphys}
m_{N}^{phys}\simeq 939\,\mathrm{MeV},\,m_{\Lambda}^{phys}\simeq 1116\,\mathrm{MeV},\,m_{\Sigma}^{phys}\simeq 1190\,\mathrm{MeV},\,m_{\Xi}^{phys}\simeq 1318\,\mathrm{MeV},\,
\end{align}
gives $\Delta_{DF}^{phys}\simeq 292\,\mathrm{MeV}$. Assuming that this choice for $D,F$ is roughly correct, and that the $\mathcal{O}(p^{4})$ terms are not too large, one would get $\frac{20}{3}b_{D}-b_{F}\sim 0.65/\mathrm{GeV}$. It is known, however, that the BChPT expressions for the baryon masses only converge slowly at the physical point. Moreover, the numerical values of $D$ and $F$ are also a source of uncertainty. We shall therefore require only that the estimate for this combination of LECs is valid within a $\sim50\%$ error,  
\begin{align}
\label{eq:bDFbound}
\frac{0.3}{\mathrm{GeV}}\lesssim \frac{20}{3}b_{D}-b_{F} \lesssim \frac{1}{\mathrm{GeV}}\quad\mathrm{for}\,D\approx\frac{3}{4}\,,F\approx\frac{1}{2}\,.
\end{align}
In contrast to $\Delta_{DF}$, the well-known Gell-Mann-Okubo difference \cite{GellMann:1961ky,Okubo:1961jc}
\begin{align}\label{eq:DeltaGMO}
\Delta_{GMO} := \frac{1}{4}\left(m_{\Sigma}+3m_{\Lambda}-2m_{N}-2m_{\Xi}\right) = \frac{D^{2}-3F^{2}}{96\pi F_{0}^{2}}\left(4\dot{M}_{K}^{3}-\dot{M}_{\pi}^{3}-3\dot{M}_{\eta}^{3}\right)+\mathcal{O}(p^{4})\,.
\end{align}
is free of $\mathcal{O}(p^{2})$-corrections. Employing the input values $D=\frac{3}{4},\,F=\frac{1}{2},\,F_{0}=80.3\,\mathrm{MeV}$, $\dot{M}_{\pi}=140\,\mathrm{MeV},\,\dot{M}_{K}=495\,\mathrm{MeV},\,\dot{M}_{\eta}= 566\,\mathrm{MeV}$ yields $\Delta_{GMO}=5.9\,\mathrm{MeV}+\mathcal{O}(p^{4})$, while $\Delta_{GMO}^{phys}\simeq 6.0\,\mathrm{MeV}$ with the baryon masses used above. It seems that the higher-order corrections in this combination are very small: the effect is almost entirely given by the leading one-loop correction. The fact that $\Delta_{GMO}$ is so small numerically could be traced back to the fact that it is of order $(\delta m_{\ell})^{2}$ in the symmetry-breaking variable.\\Fitting the $p^3$ - expressions to the experimental baryon masses, with $D=\frac{3}{4},\,F=\frac{1}{2},m_{0}^{\mathrm{eff}}:=m_{0}-6b_{0}X_{\pi}^{2}$ leads to the results presented in the table below. We vary the input parameter $F_{0}$ appearing in the $\mathcal{O}(q^{3})$-correction in a wide range to get a first impression of the uncertainty in these results.
\begin{table}
\begin{ruledtabular}
\begin{center}
\begin{tabular}{c c c c}
$F_{0}$ (MeV) & $m_{0}^{\mathrm{eff}}$ (GeV) & $b_{D}$ $(\mathrm{GeV}^{-1})$ & $b_{F}$ $(\mathrm{GeV}^{-1})$\\
\hline
70 & 2.279 & -0.039 & -0.912 \\
\hline
80 & 2.024 & -0.015 & -0.747 \\
\hline
112  & 1.617 & 0.024 & -0.484 \\
\hline
140  & 1.465 & 0.038 & -0.386  \\
\end{tabular}
\end{center} 
\end{ruledtabular}
\end{table}
The convergence of the expansion of the baryon masses using these parameters is poor, however: For example, the third order correction to the nucleon mass is $-423\,\mathrm{MeV}$ for $F_{0}=80.3\,\mathrm{MeV}$. This is a very large correction, even if one expects a generic suppression factor of $\sim\frac{M_{K}}{4\pi F_{0}}\sim 0.5$ for each additional chiral order. We infer from these results that a reliable determination of BChPT low-energy constants at (strange) quark masses higher than, or equal to, the value at the physical point is not feasible, due to much enhanced non-analytic loop corrections in this mass region. On the other hand, given that the above fit results can be regarded as a meaningful first approximation, we should at least expect the following pattern to be reproduced also in higher-order calculations:
\begin{enumerate}
\item $1\,\mathrm{GeV}\lesssim m_{0}^{\mathrm{eff}}< 3\,\mathrm{GeV}$,
\item $b_{F}$ is negative, $-1\,\mathrm{GeV}^{-1}\lesssim b_{F}< 0\,\mathrm{GeV}^{-1}$,
\item $b_{D}$ is of significantly smaller magnitude than $b_{F}$.
\end{enumerate}
\subsection*{Sigma term at third chiral order}
The third-order formula for the pion-nucleon sigma term reads (see also \cite{Bernard:1993nj}):
\begin{align}\label{eq:sigpin}
\sigma_{\pi N}(0) = -2\dot{M}_{\pi}^{2}(2b_{0}+b_{D}+b_{F}) - \frac{\dot{M}_{\pi}^{2}}{64\pi F_{0}^{2}}(4\alpha_{N}^{\pi}\dot{M}_{\pi}+2\alpha_{N}^{K}\dot{M}_{K}+\frac{4}{3}\alpha_{N}^{\eta}\dot{M}_{\eta})\,.
\end{align}
where 
\begin{align}
\alpha_{N}^{\pi}=\frac{9}{4}(D+F)^{2}\,,\quad \alpha_{N}^{K}=\frac{1}{2}(5D^{2}-6DF+9F^{2})\,,\quad \alpha_{N}^{\eta}=\frac{1}{4}(D-3F)^{2}\,.
\end{align}
Numerically, the third order correction gives a contribution of
\begin{align}
- \dot{M}_{\pi}^{2}\left(\frac{\dot{M}_{K}}{4\pi F_{0}}\right)\left(\frac{5.45}{\mathrm{GeV}}\right) \sim - 50 \,\mathrm{MeV}
\end{align}
at the physical point, to a positive quantity that is of the order $50 \,\mathrm{MeV}$ ! This again casts strong doubts on the applicability of BChPT at the physical point to obtain reasonable predictions for sigma terms, without further input.
\section{Fourth order contributions}
\label{sec:q4}
\begin{figure}[h]
\centering
\includegraphics[width=5.5cm]{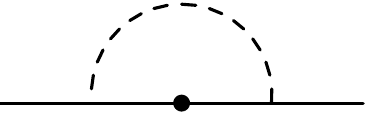}~~
\includegraphics[width=5.5cm]{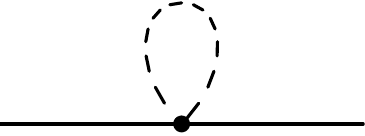}
\caption{Loop graphs giving contributions of fourth chiral order.}
\label{fig:q4loops}
\end{figure}

At fourth chiral order, additional loop graphs have to be calculated. First, one has a graph of the same topology as the graph of Fig.~\ref{fig:q3loop}, but with an additional quark mass insertion from the second order Lagrangian, proportional to $b_{0},b_{D,F}$. This graph is shown on the l.h.s. of Fig.~\ref{fig:q4loops}. In addition, there are tadpole-type contributions with vertices from the second order Lagrangian, accompanied by $b_{0},b_{D,F}$ and also eight new constants $b_{i}$,\,$i=1\ldots4,8\ldots11$. There are also seven new contact terms $\sim d_{i}$ contributing to the masses at fourth order, they absorb divergences of the infrared-regularized loop integrals. The self-energy $\Sigma^{(n)}(\slashed{p})$ of $n$-th chiral order is then calculated directly from those graphs, and gives rise to the mass shifts $\Delta m_{B}^{(n')}$ at the different orders:
\begin{align}
\begin{split}
\Delta m_{B}^{(1)} &= 0\,,\quad \Delta m_{B}^{(2)}=\Sigma^{(2)}\,,\quad \Delta m_{B}^{(3)}=\Sigma^{(3)}(\slashed{p}=m_{0})\,,\\ 
\Delta m_{B}^{(4)} &= \Sigma^{(4)}(\slashed{p}=m_{0})+\Sigma^{(2)}\frac{\partial\Sigma^{(3)}(\slashed{p})}{\partial\slashed{p}}\big|_{\slashed{p}=m_{0}}\,.\label{eq:massshifts}
\end{split}
\end{align}
The calculation of the first graph in Fig.~\ref{fig:q4loops} is straightforward and does not introduce new free parameters, so we will not discuss it further here (see e.g. Ref.~\cite{Frink:2004ic}). In the following, we are concerned with the contact terms and the tadpole graphs, which involve new unknown couplings not present in a simple leading one-loop calculation.  These new parameters can be grouped in two sets: one which will simply represent new free parameters to be determined in the fits, and one which will be fixed using input from other sources. Let us discuss both sets in turn.
\subsection*{Contact terms}
We use the fourth order Lagrangian of Frink and Mei{\ss}ner \cite{Frink:2004ic},
\begin{align}\label{eq:L4}
\begin{split}
\mathcal{L}^{(4)}_{\phi B} &= d_{1}\langle\bar{B}\lbrack\chi_{+},\lbrack\chi_{+},B\rbrack\rbrack\rangle + d_{2}\langle\bar{B}\lbrack\chi_{+},\lbrace\chi_{+},B\rbrace\rbrack\rangle + d_{3}\langle\bar{B}\lbrace\chi_{+},\lbrace\chi_{+},B\rbrace\rbrace\rangle\\
&\quad+ d_{4}\langle\bar{B}\chi_{+}\rangle\langle\chi_{+}B\rangle + d_{5}\langle\bar{B}\lbrack\chi_{+},B\rbrack\rangle\langle\chi_{+}\rangle + d_{6}\langle\bar{B}B\rangle\langle\chi_{+}\rangle\langle\chi_{+}\rangle + d_{7}\langle\bar{B}B\rangle\langle\chi_{+}^{2}\rangle\,. 
\end{split}
\end{align}
The contact terms give the following contributions to the baryon masses:
\begin{align}
\begin{split}
m_{N}^{(ct)} &= m_{\star}^{(ct)} + B_{0}\delta m_{\ell}\left(4(b_{D}-3b_{F})-16B_{0}\bar{m}(6d_{2}-4d_{3}+9d_{5})\right)\\ 
&\quad-  16(B_{0}\delta m_{\ell})^{2}\left(9d_{1}-3d_{2}+d_{3}+6d_{7}\right)\,,\\
m_{\Lambda}^{(ct)} &= m_{\star}^{(ct)} + 8B_{0}\delta m_{\ell}\left(b_{D}+16B_{0}\bar{m}d_{3}\right) - 96(B_{0}\delta m_{\ell})^{2}\left(2d_{3}+d_{4}+d_{7}\right)\,,\\
m_{\Sigma}^{(ct)}  &= m_{\star}^{(ct)} - 8B_{0}\delta m_{\ell}\left(b_{D}+16B_{0}\bar{m}d_{3}\right) - 16(B_{0}\delta m_{\ell})^{2}\left(4d_{3}+6d_{7}\right)\,,\\
m_{\Xi}^{(ct)} &= m_{\star}^{(ct)} + B_{0}\delta m_{\ell}\left(4(b_{D}+3b_{F})+16B_{0}\bar{m}(6d_{2}+4d_{3}+9d_{5})\right)\\ 
&\quad-  16(B_{0}\delta m_{\ell})^{2}\left(9d_{1}+3d_{2}+d_{3}+6d_{7}\right)\,,\label{eq:mBct}
\end{split}
\end{align}
where $m_{\star}^{(ct)}=-4B_{0}\bar{m}\left(3b_{0}+2b_{D}\right)-16(B_{0}\bar{m})^{2}\left(4d_{3}+9d_{6}+3d_{7}\right)\,$.\\
We also recall that the combination $X_{N}$ has no linear symmetry-breaking terms. In particular,
\begin{align}
X_{N}^{(ct)} = m_{\star}^{(ct)} - 32(B_{0}\delta m_{\ell})^{2}\left(3d_{1}+d_{3}+3d_{7}\right)\,.
\end{align}
In the contact term contributions linear in the symmetry-breaking, only the combinations $\tilde{b}_{D}:=b_{D}+16B_{0}\bar{m}d_{3}$ and $\tilde{b}_{F}:=b_{F}+4B_{0}\bar{m}(2d_{2}+3d_{5})$ enter. While $d_{2,3}$ also enter the terms quadratic in $\delta m_{\ell}$, we can not determine $d_{5}$ from the fan plots at the order we are working (the tadpole contribution involving $b_{F}$ is already of fourth order in the chiral counting). Moreover, the LECs $b_{0}$ and $d_{6}$ only enter in the combination
\begin{align}
m_{\star}=m_{0}-12B_{0}\bar{m}\left(b_{0}+12B_{0}\bar{m}d_{6}\right)+\ldots\,,
\end{align}
so these two parameters can also not be determined individually from the fan plots alone (again, the shift proportional to $d_{6}$ would be of higher chiral order in the tadpole graph involving $b_{0}$). For a later application, we also give the contact term contribution to the pion-nucleon sigma term. It is computed here employing the Hellmann-Feynman-theorem,
\begin{align}\label{eq:sigPiNcont1}
\begin{split}
\sigma_{\pi N}^{(ct)}(0) &=  m_{\ell}\frac{\partial m_{N}^{(ct)}}{\partial m_{\ell}}=m_{\ell}\frac{\partial}{\partial m_{\ell}}\left(m_{\star}^{(ct)}+(B_{0}\delta m_{\ell})m_{N}^{(1,ct)}+(B_{0}\delta m_{\ell})^{2}m_{N}^{(2,ct)}\right)\\
&= m_{\ell}\frac{2}{3}\frac{\partial m_{\star}^{(ct)}}{\partial \bar{m}} + \frac{1}{3}B_{0}m_{\ell}m_{N}^{(1,ct)}+(B_{0}\delta m_{\ell})\frac{2}{3}\left(m_{\ell}\frac{\partial m_{N}^{(1,ct)}}{\partial\bar{m}}+B_{0}m_{\ell}m_{N}^{(2,ct)}\right)\,. 
\end{split}
\end{align}
Inserting the expressions from eq.~(\ref{eq:mBct}), we find
\begin{align}\label{eq:sigPiNcont2}
\begin{split}
\sigma_{\pi N}^{(ct)}(0) &= -2B_{0}m_{\ell}\biggl(4b_{0}+2b_{D}+2b_{F}+\frac{32}{3}B_{0}\bar{m}\left(4d_{3}+9d_{6}+3d_{7}\right)+\frac{8}{3}B_{0}\bar{m}\left(6d_{2}-4d_{3}+9d_{5}\right)\\ 
&\quad+ 16(B_{0}\delta m_{\ell})\left(3d_{1} + d_{2} - d_{3} + 3d_{5} + 2d_{7}\right)\biggr)\,. 
\end{split}
\end{align}
\subsection*{Tadpole graphs}
The second order Lagrangian we employ here differs slightly from the one given in \cite{Frink:2004ic} because since then one further term could be eliminated \cite{Frink:2011pc}.
The Lagrangian is then given by
\begin{align}\label{eq:L2}
\begin{split}
\mathcal{L}^{(2)}_{\phi B}&= b_{D/F} \langle\overline B\left[\chi_+,B\right]_\pm\rangle+b_0 \langle\overline B B\rangle \langle\chi_+\rangle\\
&\quad+b_{1/2} \langle\overline B  \left[u_\mu,\left[u^\mu,B\right]_\mp\right]\rangle +b_3 \langle\overline B \left\{ u_\mu,\left\{ u^\mu,B\right\}\right\}\rangle +b_4 \langle\overline B  B\rangle \langle u_\mu u^\mu\rangle \\
&\quad+i\Big(b_{5/6}  \langle\overline B \sigma^{\mu\nu}\left[\left[u_\mu,u_\nu\right], B\right]_\mp\rangle +b_7 \langle\overline B \sigma^{\mu\nu}u_\mu\rangle  \langle u_\nu B\rangle\Big)\\
&\quad+ \frac{i\,b_{8/9}}{2m_0}\Big( \langle\overline B \gamma^\mu\left[u_\mu,\left[u_\nu,\left[D^\nu, B\right]\right]_\mp\right]\rangle+\langle\overline B \gamma^\mu\big[D_\nu,\left[u^\nu,\left[u_\mu,B\right]\right]_\mp\big]\rangle\Big)\\
&\quad+\frac{i\,b_{10}}{2m_0}\Big( \langle\overline B \gamma^\mu\left\{ u_\mu,\left\{ u_\nu,\left[D^\nu,B\right]\right\}\right\}\rangle+\langle\overline B\gamma^\mu\left[D_\nu,\left\{ u^\nu,
\left\{ u_\mu,B\right\}\right\}\right]\rangle\Big)\\
&\quad+\frac{i\,b_{11}}{2m_0}\Big( 2\langle\overline B \gamma^\mu \left[D_\nu,B\right]\rangle \langle u_\mu u^\nu\rangle + \langle\overline B \gamma^\mu B\rangle \langle\left[D_\nu,u_\mu\right]u^\nu + u_\mu \left[D_\nu,u^\nu\right]\rangle   \Big)~. 
\end{split}
\end{align}
This form has also been used in \cite{Bruns:2010sv,Mai:2012wy}, and we shall also take the numerical values of the LECs $b_{1-4},\,b_{8-11}$ from the latter references. 
Note, however, that in those references, the chiral potential derived from the above Lagrangian is iterated to infinite order in a coupled-channel Bethe-Salpeter equation, introducing a model-dependent uncertainty not controlled within strict ChPT. To estimate the impact of this uncertainty on our results, we perform all our fits with three different parameter sets from meson-baryon scattering (set 1-3). In addition, we also add one parameter set where we take all $b_{1-11}$ to vanish (set 4). The four parameter sets are collected in the table below (in $\mathrm{GeV}^{-1}$).
\begin{table}[h]
\caption{Four parameter sets for the low-energy coefficients $b_{1-4},b_{8-11}$.}
\label{tab:bi}
\begin{center}
\begin{ruledtabular}
\begin{tabular}{c c c c c c c c c c}
 set & $b_{1}$ & $b_{2}$ & $b_{3}$ & $b_{4}$ & $b_{8}$ & $b_{9}$ & $b_{10}$ & $b_{11}$ & Ref. \\
\hline
 1 &  -0.082 & -0.118 & -1.890 & -0.215 & 0.609 & -0.633 & 1.920 & -0.919 & \cite{Bruns:2010sv} \\ 
\hline
 2 & -0.126 & -0.139 & -2.227 & -0.288 & 0.610 & -0.677 & 2.027 & -0.847 & \cite{Mai:2012wy} \\
\hline
 3 & -0.014 & -0.207 & -1.063 & -1.312 & 0.272 & -0.483 & 1.054 & 0.328 & \cite{Mai:2012wy} \\
\hline
 4 & 0 & 0 & 0 & 0 & 0 & 0 & 0 & 0 & - \\
\end{tabular}
\end{ruledtabular}
\end{center}
\end{table}
$b_{5-7}$ do not contribute to the baryon masses at one-loop order. We shall not write out the full tadpole contributions to the baryon masses. The corresponding contributions to the terms linear in the symmetry-breaking $\delta m_{\ell}$ and to the baryon mass in the symmetry limit, $m_{\star}$, can be read off from eqs.~(\ref{eq:tadlinf}) and (\ref{eq:mstar}).
\subsection*{Renormalization}
The infrared regularized loop integrals still contain UV divergences as $d\rightarrow 4$, proportional to
\begin{align}
L:=\frac{\mu^{d-4}}{16\pi^{2}}\left(\frac{1}{d-4}-\frac{1}{2}\left[\ln(4\pi)+\Gamma'(1)+1\right]\right)\,.
\end{align}
The couplings $b_{0,D,F}$ do not have to absorb divergences from the infrared regularized loop graphs. At third order, there are no counterterms contributing to the baryon masses, and therefore also no divergences. The divergences of the infrared singular parts of the loop integrals at fourth order can be absorbed by the following renormalization prescription for the counterterms $d_{i}$:
\begin{align}
d_{i}= \gamma_{i}^{\mathrm{IR}}L + d_{i}^{(r)}(\mu)\,,
\end{align}
with
\begin{align}
\begin{split}
\gamma_{1}^{\mathrm{IR}} &= \frac{1}{72F_{0}^{2}}\biggl(b_{D}(14+69D^2 + 81F^2) + 162b_{F}DF + \frac{D^2-3F^2}{m_{0}} - (12b_{1}-4b_{3}+3b_{8}-b_{10})\biggr)\,,\\
\gamma_{2}^{\mathrm{IR}} &= \frac{1}{48F_{0}^{2}}\biggl(120b_{D}DF + b_{F}(4+60D^2+108F^2)+\frac{6DF}{m_{0}} + 3(4b_{2}+b_{9})\biggr)\,,\\
\gamma_{3}^{\mathrm{IR}} &= \frac{1}{48F_{0}^{2}}\biggl(6b_{D}(4+13D^2+9F^2)+108b_{F}DF + \frac{3(D^2-3F^2)}{m_{0}}-(36b_{1}+4b_{3}+9b_{8}+b_{10})\biggr)\,,\\
\gamma_{4}^{\mathrm{IR}} &= \frac{1}{72F_{0}^{2}}\biggl(-4b_{D}(11+72D^2)-\frac{9(D^2-3F^2)}{m_{0}}+108b_{1}-4b_{3}+27b_{8}-b_{10}\biggr)\,,\\
\gamma_{5}^{\mathrm{IR}} &= \frac{1}{72F_{0}^{2}}\biggl(44b_{F}-\frac{26DF}{m_{0}}-13(4b_{2}+b_{9})\biggr)\,,\\
\gamma_{6}^{\mathrm{IR}} &= \frac{1}{432F_{0}^{2}}\biggl(264b_{0}+b_{D}(132-144D^2)-\frac{35D^2+27F^2}{m_{0}}\\ 
&\quad- (108b_{1}+140b_{3}+264b_{4}+27b_{8}+35b_{10}+66b_{11})\biggr)\,,\\
\gamma_{7}^{\mathrm{IR}} &= \frac{1}{144F_{0}^{2}}\biggl(120b_{0}+b_{D}(28-24(7D^2+9F^2))-432b_{F}DF-\frac{17D^2+9F^2}{m_{0}}\\ 
&\quad- (36b_{1}+68b_{3}+120b_{4}+9b_{8}+17b_{10}+30b_{11})\biggr)\,.\label{eq:Renorm} 
\end{split}
\end{align}
We can then define scale-independent quantities $\bar{d}_{i}$ by 
\begin{align}
d_{i}^{(r)}(\mu) = \bar{d}_{i} + \frac{\gamma_{i}^{\mathrm{IR}}}{16\pi^{2}}\log\left(\frac{m_{0}}{\mu}\right)\,.\label{eq:drun}
\end{align}
\section{Analysis of fan plot data}
\label{sec:Analyse}
The main objective of this study is the analysis of the fan plot data collected in Tab.~22 and Fig.~20 of \cite{Bietenholz:2011qq}. The following ratios of baryon masses are considered,
\begin{align}\label{eq:fB}
f_{B}:=\frac{m_{B}}{X_{N}}\qquad\mathrm{for}\quad B=N,\Lambda,\Sigma,\Xi\,,
\end{align}
see also eq.~(\ref{eq:XN}) for the definition of $X_{N}$. At the point on the symmetric line where $m_{\ell}=m_{s}=\bar{m}$, we have $f_{B}=\frac{m_{\star}}{m_{\star}}=1$ and $X_{\pi}=M_{\star}=412\,\mathrm{MeV}$. The symmetry breaking is then switched on by varying $\delta m_{\ell}$ for fixed $\bar{m}$. For the data points with the greatest distance from the symmetric point, we compute $\nu=-0.692$ from eq.~(\ref{eq:nu}). The meson masses at this point are $M_{\pi}\approx 229\,\mathrm{MeV}$, $M_{K}\approx 477\,\mathrm{MeV}$ and (using eq.~(\ref{eq:MesGMO})) $M_{\eta}\approx 535\,\mathrm{MeV}$. All meson masses stay well below the experimental eta mass, in contrast to other lattice data sets (with $m_{s}$ fixed instead of $\bar{m}$), where the meson masses are usually {\em larger}\, than the meson masses in the real world. There is therefore good reason to believe that the chiral extrapolation formulae provided by three-flavor BChPT will work better when applied to the present data set than in the usual applications, though meson masses of $\sim 400\,\mathrm{MeV}$ are probably still too high to assure a proper convergence behavior of the chiral expansions in general. Another indication of a better behavior of the quantities considered here is the observation (made in \cite{Bietenholz:2011qq}) that finite-volume effects, which are mainly generated by the chiral loops, tend to cancel out in the baryon mass ratios $f_{B}$. Here we only remark that finite volume effects might still be non-negligible for the data in question, and deserve further study. It is also interesting to note that, employing large-$N_{c}$ arguments and Heavy-Baryon ChPT, it was shown in \cite{WalkerLoud:2011ab} that the poor convergence behavior of the three-flavor chiral extrapolations could be traced back to the flavor-singlet sector, considering in particular the mass relation called $R_{1}$ in the latter reference. This again suggests that it might be a good idea to factor out a convenient flavor-singlet quantity (in our case, $X_{N}$) as is done in the baryon mass ratios relevant for the fan plots. In this section, we will see that certain combinations of low-energy constants can already be extracted quite reliably from the fan plot data for the ratios $f_{B}$, namely, the LEC combinations which parameterize the leading symmetry-breaking contributions to the baryon masses. We consider this as a first and necessary step toward a theoretically controlled application of BChPT formulae to lattice data. We also make an attempt to determine the remaining LECs occuring in the baryon mass extrapolations, however, the constants parameterizing the singlet contributions can only be roughly estimated.

Though we shall use the full one-loop BChPT expressions for the baryon masses in the ratios $f_{B}$ in the end, it is instructive to consider the expansion of the latter ratios in the variable $\delta m_{\ell}$ parameterizing the symmetry breaking. This will make clear which subsets and combinations of LECs can be more accurately determined than before, and what the qualifications and sources of uncertainties for the various determinations are.
\subsection*{Expansion of mass ratios $f_{B}$}
Employing the full one-loop ChPT calculation of the baryon masses, we are sensitive only up to terms $(\delta m_{\ell})^{k}\bar{m}^{2-k}$, while higher order terms will be modified by terms on the two-loop level. Therefore, the following equations are to be understood as resulting from a double expansion: the chiral expansion, on the one hand, and the expansion in $\delta m_{\ell}$, on the other hand. One finds
\begin{align}\label{eq:fBexp}
f_{B} = 1+(B_{0}\delta m_{\ell})f_{B}^{(1)} + (B_{0}\delta m_{\ell})^{2}f_{B}^{(2)}+\mathcal{O}((\delta m_{\ell})^{3})\,,
\end{align}
where
\begin{align}
\begin{split}
f_{N}^{(1)} &= \frac{4(b_{D}-3b_{F})}{m_{0}} - \sqrt{2B_{0}\bar{m}}\frac{3(D^{2}+10DF-3F^{2})}{32m_{0}\pi F_{0}^{2}} \\ 
&\quad+ (16B_{0}\bar{m})\left(\frac{(3b_{0}+2b_{D})(b_{D}-3b_{F})}{m_{0}^{2}} - \frac{6d_{2}-4d_{3}+9d_{5}}{m_{0}} +\ldots\right)+\mathcal{O}(\bar{m}^{3/2})\,,
\end{split}\\
\begin{split}
f_{\Lambda}^{(1)} &= \frac{8b_{D}}{m_{0}} - \sqrt{2B_{0}\bar{m}}\frac{3(D^{2}-3F^{2})}{16m_{0}\pi F_{0}^{2}} \\ 
&\quad+ (16B_{0}\bar{m})\left(\frac{2(3b_{0}+2b_{D})b_{D}}{m_{0}^{2}} + \frac{8d_{3}}{m_{0}} +\ldots\right)+\mathcal{O}(\bar{m}^{3/2})\,, 
\end{split}\\
\begin{split}
f_{\Sigma}^{(1)} &= -\frac{8b_{D}}{m_{0}}+ \sqrt{2B_{0}\bar{m}}\frac{3(D^{2}-3F^{2})}{16m_{0}\pi F_{0}^{2}} \\  
&\quad- (16B_{0}\bar{m})\left(\frac{2(3b_{0}+2b_{D})b_{D}}{m_{0}^{2}} + \frac{8d_{3}}{m_{0}} +\ldots\right)+\mathcal{O}(\bar{m}^{3/2})\,, 
\end{split}\\
\begin{split}
f_{\Xi}^{(1)} &= \frac{4(b_{D}+3b_{F})}{m_{0}} - \sqrt{2B_{0}\bar{m}}\frac{3(D^{2}-10DF-3F^{2})}{32m_{0}\pi F_{0}^{2}} \\ 
&\quad+ (16B_{0}\bar{m})\left(\frac{(3b_{0}+2b_{D})(b_{D}+3b_{F})}{m_{0}^{2}} + \frac{6d_{2}+4d_{3}+9d_{5}}{m_{0}} +\ldots\right)+\mathcal{O}(\bar{m}^{3/2})\,, 
\end{split}
\end{align}
and
\begin{align}
f_{N}^{(2)} &= -\frac{D^{2}+18DF-3F^{2}}{128\sqrt{2B_{0}\bar{m}}m_{0}\pi F_{0}^{2}} -16\frac{3d_{1}-3d_{2}-d_{3}}{m_{0}}+\ldots + \mathcal{O}(\bar{m}^{1/2})\,,\\
f_{\Lambda}^{(2)} &= -\frac{10(D^{2}-3F^{2})}{128\sqrt{2B_{0}\bar{m}}m_{0}\pi F_{0}^{2}}+32\frac{3d_{1}-5d_{3}-3d_{4}}{m_{0}}+\ldots + \mathcal{O}(\bar{m}^{1/2})\,,\\
f_{\Sigma}^{(2)}  &= \frac{2(D^{2}-3F^{2})}{128\sqrt{2B_{0}\bar{m}}m_{0}\pi F_{0}^{2}}+32\frac{3d_{1}-d_{3}}{m_{0}}+\ldots + \mathcal{O}(\bar{m}^{1/2})\,,\\
f_{\Xi}^{(2)} &= -\frac{D^{2}-18DF-3F^{2}}{128\sqrt{2B_{0}\bar{m}}m_{0}\pi F_{0}^{2}}-16\frac{3d_{1}+3d_{2}-d_{3}}{m_{0}}+\ldots + \mathcal{O}(\bar{m}^{1/2})\,.\label{eq:fB2}
\end{align}
The dots stand for corrections from loop graphs of fourth chiral order, which we do not display explicitly here. Let us concentrate first on the contribution from the contact terms. We make the following observations:
\begin{enumerate}
\item{$f_{\Sigma}^{(1)} = -f_{\Lambda}^{(1)}$,}
\item{$f_{N}^{(n)}+f_{\Sigma}^{(n)}+f_{\Xi}^{(n)}=0$,}
\item{the baryon mass $m_{0}$ only appears in combination with low-energy constants, in the ratios $D/\sqrt{m_{0}},\,F/\sqrt{m_{0}}$, $b_{i}/m_{0},\,d_{i}/m_{0}\,$,}
\item{the LEC $d_{7}$ does not appear at this order in the double expansion of the ratios.}
\end{enumerate}
The last two items would lead to problems when only using the fan plots and ratios $f_{B}$ in the fitting procedure: First, since terms of order $(\delta m_{\ell})^{3}$ and higher are expected to be small in most parts of the fan plots, it is very likely that a stable determination of $d_{7}$ would be difficult. Second, we would like to use the values of $D,F,b_{1-4},b_{8-11}$ directly as input, and the baryon mass in the chiral limit, $m_{0}$, is not accurately known. What is more or less accurately known, however, is the baryon mass at the symmetric point, 
\begin{align}
m_{\star}=m_{0}-4B_{0}\bar{m}(3b_{0}+2b_{D})+\mathcal{O}(\bar{m}^{3/2})\,,
\end{align}
for reasons discussed in the next subsection. We can eliminate $m_{0}$ in favor of $m_{\star}$. Moreover, we can replace $F_{0}$ by $F_{\star}$ (see eq.~(\ref{eq:Fstar})) at the order we are working. These replacements are rather natural when expanding in the symmetry breaking around the point $m_{\ell}=m_{s}=\bar{m}$. Doing this, the previous expansions read after this slight reordering (in terms of $\tilde{b}_{D,F}$ defined above):
\begin{align}
\begin{split}
f_{N}^{(1)} &= \frac{4(\tilde{b}_{D}-3\tilde{b}_{F})}{m_{\star}} - \sqrt{2B_{0}\bar{m}}\frac{3(D^{2}+10DF-3F^{2})}{32m_{\star}\pi F_{\star}^{2}}\\ 
&\quad- \frac{4B_{0}\bar{m}}{(4\pi F_{\star})^{2}}\bigg[\frac{\tilde{b}_{D}-3\tilde{b}_{F}}{m_{\star}}\left(\frac{5}{3}+8\log\left(\frac{\sqrt{2B_{0}\bar{m}}}{\mu}\right)\right)\\ 
&\quad+ \frac{2\tilde{b}_{D}}{m_{\star}}(13D^{2}-30DF+9F^{2})\left(\frac{1}{3}+\log\left(\frac{\sqrt{2B_{0}\bar{m}}}{\mu}\right)\right)\\ 
&\quad- \frac{6\tilde{b}_{F}}{m_{\star}}(5D^{2}-6DF+9F^{2})\left(\frac{1}{3}+\log\left(\frac{\sqrt{2B_{0}\bar{m}}}{\mu}\right)\right)\\ 
&\quad+ \frac{D^{2}+10DF-3F^{2}}{m_{\star}^{2}}\left(\frac{3}{4}+\log\left(\frac{\sqrt{2B_{0}\bar{m}}}{\mu}\right)\right)\bigg]  + \mathcal{T}^{(1)}_{N} +\mathcal{O}(\bar{m}^{3/2})\,,
\end{split}\\
\begin{split}
f_{\Lambda}^{(1)} &= \frac{8\tilde{b}_{D}}{m_{\star}} - \sqrt{2B_{0}\bar{m}}\frac{3(D^{2}-3F^{2})}{16m_{\star}\pi F_{\star}^{2}}  - \frac{4B_{0}\bar{m}}{(4\pi F_{\star})^{2}}\bigg[\frac{2\tilde{b}_{D}}{m_{\star}}\left(\frac{5}{3}+8\log\left(\frac{\sqrt{2B_{0}\bar{m}}}{\mu}\right)\right)\\ 
&\quad+ \frac{4\tilde{b}_{D}}{m_{\star}}(13D^2 + 9F^2)\left(\frac{1}{3}+\log\left(\frac{\sqrt{2B_{0}\bar{m}}}{\mu}\right)\right) + \frac{72\tilde{b}_{F}}{m_{\star}}DF\left(\frac{1}{3}+\log\left(\frac{\sqrt{2B_{0}\bar{m}}}{\mu}\right)\right)\\ 
&\quad+ \frac{2(D^2 - 3F^2)}{m_{\star}^{2}}\left(\frac{3}{4}+\log\left(\frac{\sqrt{2B_{0}\bar{m}}}{\mu}\right)\right)\bigg]  + \mathcal{T}^{(1)}_{\Lambda} +\mathcal{O}(\bar{m}^{3/2})\,,
\end{split}\\
\begin{split}
f_{\Sigma}^{(1)} &= -f_{\Lambda}^{(1)}\,, 
\end{split}\\
\begin{split}
f_{\Xi}^{(1)} &= -(f_{N}^{(1)}+f_{\Sigma}^{(1)})\,,\label{eq:feins} 
\end{split}
\end{align}
while the formulae for $f_{B}^{(2)}$ just change in that $m_{0}\rightarrow m_{\star}$ at the chiral order we are working. $\mathcal{T}^{(1)}_{B}$ stand for contributions from tadpole graphs proportional to $b_{1-4},b_{8-11}$. These are given below. We see that from the derivative of the ratios at the symmetric point, we can determine two parameters only, due to the symmetry constraints in (i) and (ii), namely $\tilde{b}_{D,F}$. From the tadpole contributions to the $f^{(1)}_{B}$,
\begin{align}
\begin{split}
\mathcal{T}^{(1)}_{N} &= \frac{4B_{0}\bar{m}}{3(4\pi F_{\star})^{2}}\bigg[\frac{9b_{1}-15b_{2}+b_{3}}{m_{\star}}\left(1+4\log\left(\frac{\sqrt{2B_{0}\bar{m}}}{\mu}\right)\right) + \frac{9b_{8}-15b_{9}+b_{10}}{m_{\star}}\log\left(\frac{\sqrt{2B_{0}\bar{m}}}{\mu}\right)\bigg]\,,\\
\mathcal{T}^{(1)}_{\Lambda} &= \frac{8B_{0}\bar{m}}{3(4\pi F_{\star})^{2}}\bigg[\frac{9b_{1}+b_{3}}{m_{\star}}\left(1+4\log\left(\frac{\sqrt{2B_{0}\bar{m}}}{\mu}\right)\right) + \frac{9b_{8}+b_{10}}{m_{\star}}\log\left(\frac{\sqrt{2B_{0}\bar{m}}}{\mu}\right)\bigg]\,,\label{eq:tadlinf}
\end{split}
\end{align}
we can see that $b_{2,9}$ enter there only for $N,\Xi$, while the combinations $9b_{1}+b_{3}$,  $9b_{8}+b_{10}$ enter for all four baryons.
\subsection*{Expansion of $X_{N}$}
The above ratios $f_{B}$ are all normalized to $X_{N}:=\frac{1}{3}(m_{N}+m_{\Sigma}+m_{\Xi})$. This combination was shown in \cite{Bietenholz:2011qq} to be practically constant along a trajectory with constant $\bar{m}$, and thus approximately equal to $m_{\star}^{num}=1150\,\mathrm{MeV}$ for the choice of $\bar{m}$ corresponding to the experimental hadron masses. Subsequently, $X_{N}$ was even used to set the scale in the simulations leading to the fan plot for the baryon masses. Employing our ChPT formulae, the flat behavior of $X_{N}$ is not automatically guaranteed, but should be enforced in the fitting procedure. To see how this can be done, we write down the expansion of $X_{N}$ in the symmetry-breaking variable $\delta m_{\ell}$:
\begin{align}
X_{N}= m_{\star} + (B_{0}\delta m_{\ell})^{2}X_{N}^{(2)} + \mathcal{O}((\delta m_{\ell})^{3})\,,
\end{align}
with
\begin{align}
\begin{split}
m_{\star} &= m_{0}-4B_{0}\bar{m}(3b_{0}+2b_{D}) - (2B_{0}\bar{m})^{3/2}\frac{(5D^{2}+9F^{2})}{24\pi F_{\star}^{2}}\\
&\quad+ \frac{(2B_{0}\bar{m})^{2}}{3(4\pi F_{\star})^{2}}\bigg[32(3b_{0}+2b_{D})\log\left(\frac{\sqrt{2B_{0}\bar{m}}}{\mu}\right) - 8(9b_{1}+7b_{3}+12b_{4})\log\left(\frac{\sqrt{2B_{0}\bar{m}}}{\mu}\right)\\
&\quad- 2(9b_{8}+7b_{10}+12b_{11})\left(\log\left(\frac{\sqrt{2B_{0}\bar{m}}}{\mu}\right) - \frac{1}{4}\right)  -12(4\pi F_{\star})^{2}\left(4d_{3}+9d_{6}+3d_{7}\right)\\ 
&\quad-  \frac{(5D^{2}+9F^{2})}{m_{0}}\left(1+2\log\left(\frac{\sqrt{2B_{0}\bar{m}}}{\mu}\right)\right)\bigg]  + \mathcal{O}(\bar{m}^{5/2})\,,\label{eq:mstar} 
\end{split}
\end{align}
and 
\begin{align}
\begin{split}
X_{N}^{(2)} &=  -\frac{3(D^{2}+2F^{2})}{16\pi F_{\star}^{2}\sqrt{2B_{0}\bar{m}}} - 32\left(3d_{1}+d_{3}+3d_{7}\right)+\frac{5b_{0}}{\pi^{2}F_{\star}^{2}}\left(\frac{3}{4}+\log\left(\frac{\sqrt{2B_{0}\bar{m}}}{\mu}\right)\right)\\ 
&\quad+ \frac{b_{D}}{6\pi^{2}F_{\star}^{2}}\left(2D^{2}\left(5+6\log\left(\frac{\sqrt{2B_{0}\bar{m}}}{\mu}\right)\right)+15+20\log\left(\frac{\sqrt{2B_{0}\bar{m}}}{\mu}\right)\right)\\
&\quad- \frac{\left(12b_{1}+8b_{3}+15b_{4}\right)}{12\pi^{2}F_{\star}^{2}}\left(3+4\log\left(\frac{\sqrt{2B_{0}\bar{m}}}{\mu}\right)\right)\\
&\quad- \frac{\left(12b_{8}+8b_{10}+15b_{11}\right)}{24\pi^{2}F_{\star}^{2}}\left(1+2\log\left(\frac{\sqrt{2B_{0}\bar{m}}}{\mu}\right)\right)\\
&\quad- \frac{D^{2}+2F^{2}}{2m_{0}\pi^{2}F_{\star}^{2}}\left(\frac{5}{4}+\log\left(\frac{\sqrt{2B_{0}\bar{m}}}{\mu}\right)\right) + \mathcal{O}(\bar{m}^{1/2})\,.\label{eq:XN2} 
\end{split}
\end{align}
Here, the LEC $d_{7}$ shows up at order $\bar{m}^{2}$ and $(\delta m_{\ell})^{2}$, respectively. The baryon mass in the chiral limit,  $m_{0}$, and the combinations of LECs appearing here will be determined in the next section from the behavior of $m_{\star}(\bar{m})$, for which some lattice data points with $300\,\mathrm{MeV}<M_{\star}<500\,\mathrm{MeV}$ exist. We repeat that we dismiss any lattice data involving meson masses much above $500\,\mathrm{MeV}$. 
\subsection*{Linear fits to the fan plots}
As a first step, we perform a linearized fit to the fan plot data only. Only the LEC-combinations that enter terms linear in the symmetry breaking can be safely determined from such fits.
As shown above, in a full one-loop calculation in ChPT, these are the combinations $\tilde{b}_{D,F}$. As an additional constraint, we impose that the mass combination $X_{N}$ does not deviate much from the physical value $\sim 1150\,\mathrm{MeV}$. Approximating 
\begin{align}
\begin{split}
f_{B} &\approx 1+\nu f_{B}^{(1,\nu)}\,,\\
X_{N} &\approx m_{\star}+\nu^{2}X_{N}^{(2,\nu)}\,,\label{eq:approxf} 
\end{split}
\end{align}
using eqs.~(\ref{eq:feins}) and (\ref{eq:nuRel}), and truncating the coefficients after $\mathcal{O}((2B_{0}\bar{m})^{2})$, we can determine the following four combinations of LECs from such fits:
\begin{align}
\begin{split}
\tilde{b}_{D} &= b_{D}+16B_{0}\bar{m}d_{3}\,,\qquad \tilde{b}_{F} = b_{F}+4B_{0}\bar{m}(2d_{2}+3d_{5})\equiv b'_{F}+8B_{0}\bar{m}d_{2}\,,\\
\tilde{b}_{0} &= b_{0}-2B_{0}\bar{m}\left(\frac{8}{3}d_{3}-6d_{6}-2d_{7}\right)\,,\qquad \tilde{d}_{7} = d_{1}+\frac{1}{3}d_{3}+d_{7}\,.\label{eq:deflinpar} 
\end{split}
\end{align}
The loop functions are evaluated at a scale $\mu=1150\,\mathrm{MeV}$. We have checked that, given the running of the coupling constants with $\mu$ from eq.~(\ref{eq:drun}), the masses are independent of the choice of scale when truncating after $\mathcal{O}(p^{4})$.\\ In our numerical analysis, we use $2B_{0}\bar{m}=(420\,\mathrm{MeV})^{2}$ in order to fix the meson mass at the symmetric point, $M_{\star}$, to $412\,\mathrm{MeV}$ for the set of meson LECs from Tab.~\ref{tab:Li}. $m_{\star}$ is set to $m_{\star}^{num}=1150\,\mathrm{MeV}$ in the expressions for $f_{B}^{(1,\nu)}$, eliminating the dependence of these quantities on $m_{0}$ (compare eq.~(\ref{eq:feins})), while $F_{\star}$ is varied in the range $(112\pm30)\,\mathrm{MeV}$ to get an estimate for the error of the determined combinations (one finds that the results are not very sensitive to the choice of $2B_{0}\bar{m}$). The results of these fits are given in the tables below, for some given values of $m_{0}$ and $F_{\star}$, and for all four parameter sets of $b_{1-11}$ from Tab.~\ref{tab:bi}.
\begin{table}[h]
\begin{ruledtabular}
\caption{$b_{1-11}$ from set 1}
\begin{center}
\begin{tabular}{c c c c c c}
$F_{\star}$ (MeV) & $m_{0}$ (MeV) & $\tilde{b}_{D}$ $(\mathrm{GeV}^{-1})$ & $\tilde{b}_{F}$ $(\mathrm{GeV}^{-1})$ & $\tilde{b}_{0}$ $(\mathrm{GeV}^{-1})$ & $\tilde{d}_{7}$ $(\mathrm{GeV}^{-3})$ \\
\hline
112 & 800 & 0.078 & -0.352 & -0.888 & -0.158 \\ 
\hline
112 & 1000 & 0.078 & -0.352 & -0.764 & -0.168 \\
\hline
112 & 1200 & 0.078 & -0.352 & -0.639 & -0.180 \\
\hline
80  & 800 & 0.109 & -0.434 & -1.158 & -0.257 \\
\hline
80  & 1000 & 0.109 & -0.434 & -1.067 & -0.271 \\
\hline
80  & 1200 & 0.109 & -0.434 & -0.974 & -0.287 \\
\hline
140  & 1000 & 0.066 & -0.306 & -0.599 & -0.118 \\
\end{tabular}
\end{center}  
\end{ruledtabular}
\begin{ruledtabular}
\caption{$b_{1-11}$ from set 2}
\begin{center}
\begin{tabular}{c c c c c c}
$F_{\star}$ (MeV) & $m_{0}$ (MeV) & $\tilde{b}_{D}$ $(\mathrm{GeV}^{-1})$ & $\tilde{b}_{F}$ $(\mathrm{GeV}^{-1})$ & $\tilde{b}_{0}$ $(\mathrm{GeV}^{-1})$ & $\tilde{d}_{7}$ $(\mathrm{GeV}^{-3})$ \\
\hline
112 & 800 & 0.062 & -0.355 & -0.961 & -0.173 \\ 
\hline
112 & 1000 & 0.062 & -0.355 & -0.838 & -0.183 \\
\hline
112 & 1200 & 0.062 & -0.355 & -0.713 & -0.195 \\
\hline
80  & 800 & 0.088 & -0.437 & -1.270 & -0.277 \\
\hline
80  & 1000 & 0.088 & -0.437 & -1.179 & -0.291 \\
\hline
80  & 1200 & 0.088 & -0.437 & -1.086 & -0.307 \\
\hline
140  & 1000 & 0.054 & -0.308 & -0.652 & -0.130 \\
\end{tabular}
\end{center} 
\end{ruledtabular}
\end{table}
\begin{table}
\begin{ruledtabular}
\caption{$b_{1-11}$ from set 3}
\begin{center}
\begin{tabular}{c c c c c c}
$F_{\star}$ (MeV) & $m_{0}$ (MeV) & $\tilde{b}_{D}$ $(\mathrm{GeV}^{-1})$ & $\tilde{b}_{F}$ $(\mathrm{GeV}^{-1})$ & $\tilde{b}_{0}$ $(\mathrm{GeV}^{-1})$ & $\tilde{d}_{7}$ $(\mathrm{GeV}^{-3})$ \\
\hline
112 & 800 & 0.083 & -0.358 & -1.024 & -0.181 \\ 
\hline
112 & 1000 & 0.083 & -0.358 & -0.900 & -0.192 \\
\hline
112 & 1200 & 0.083 & -0.358 & -0.775 & -0.203 \\
\hline
80  & 800 & 0.117 & -0.443 & -1.362 & -0.289 \\
\hline
80  & 1000 & 0.117 & -0.443 & -1.271 & -0.303 \\
\hline
80  & 1200 & 0.117 & -0.443 & -1.178 & -0.319 \\
\hline
140  & 1000 & 0.069 & -0.310 & -0.697 & -0.136 \\
\end{tabular}
\end{center} 
\end{ruledtabular}
\begin{ruledtabular}
\caption{$b_{1-11}$ from set 4}
\begin{center}
\begin{tabular}{c c c c c c}
$F_{\star}$ (MeV) & $m_{0}$ (MeV) & $\tilde{b}_{D}$ $(\mathrm{GeV}^{-1})$ & $\tilde{b}_{F}$ $(\mathrm{GeV}^{-1})$ & $\tilde{b}_{0}$ $(\mathrm{GeV}^{-1})$ & $\tilde{d}_{7}$ $(\mathrm{GeV}^{-3})$ \\
\hline
112 & 800 & 0.072 & -0.305 & -0.504 & -0.091 \\ 
\hline
112 & 1000 & 0.072 & -0.305 & -0.380 & -0.101 \\
\hline
112 & 1200 & 0.072 & -0.305 & -0.255 & -0.113 \\
\hline
80  & 800 & 0.095 & -0.365 & -0.580 & -0.166 \\
\hline
80  & 1000 & 0.095 & -0.365 & -0.490 & -0.180 \\
\hline
80  & 1200 & 0.095 & -0.365 & -0.396 & -0.196 \\
\hline
140  & 1000 & 0.063 & -0.271 & -0.322 & -0.068 \\
\end{tabular}
\end{center} 
\end{ruledtabular}
\end{table}
Based on the experience with these fits, and on the expectation of higher order corrections to the approximations made in eqs.~(\ref{eq:approxf}), we can impose the following bounds on the parameter combinations in question (in appropriate units, specified in the above tables):
\begin{align}
 0.05 \leq &\,\tilde{b}_{D} \leq 0.15\,,\label{eq:bDtillin}\\
-0.50 \leq &\,\tilde{b}_{F} \leq -0.25\,,\label{eq:bFtillin}\\
-1.50 \leq &\,\tilde{b}_{0} \leq -0.20\,,\label{eq:b0tillin}\\
-0.35 \leq &\,\tilde{d}_{7} \leq -0.05\,.\label{eq:d7tillin}
\end{align}
The parameter combinations $\tilde{b}_{D,F}$ will be fixed in the full fits for every set $1-4$, while $m_{0},b_{0},d_{1-7}$ will be left free - the comparison with the values $\tilde{b}_{0},\tilde{d}_{7}$ from the tables above will only serve as a consistency check afterwards. Please note that the numerical values given in the previous four equations relate to the fixed values for $\mu$ and $2B_{0}\bar{m}$ specified above. 
\subsection*{Full one-loop fits to the fan plots}
In the next step, we use the full one-loop expressions for the baryon masses in the fit functions. We fit to the fan plot data for the baryon octet masses (Table 22 of \cite{Bietenholz:2011qq}) and the lowest three points for $X_{N}(\bar{m},\nu=0)=m_{\star}(M_{\star})$ (from Table 19 of \cite{Bietenholz:2011qq}) where $M_{\star}\leq 412\,\mathrm{MeV}$. In addition, we require that $X_{N}(\nu)\approx 1150\,\mathrm{MeV}$ for $\nu\in\lbrace -0.692, -0.558, -0.404, -0.275, -0.128, 0, 0.181\rbrace$ (i.e. at the fan plot data points), where we allow for an error of 10\% for  $X_{N}(\nu)$ at those points. $m_{\star}(M_{\star})$ is derived from eq.~(\ref{eq:mstar}) with the help of eq.~(\ref{eq:Mstar}). Throughout we eliminate $F_{0}$ in favor of $F_{\star}$ - the difference, however, formally amounts to an $\mathcal{O}(p^{5})$-effect in the baryon masses, which is beyond the order we are working in. The ratios $f_{B}$ are directly obtained by inserting the one-loop expressions for the baryon masses, without further expansion in $\nu$ or $\bar{m}$. Also, $X_{N}(\nu)$ is not truncated after $\mathcal{O}(\nu^{2})$ in the full fits. In the fan plot fitting functions, including $X_{N}(\nu)$, where the average quark mass is fixed to its physical value, we shall use the numerical values from eqs.~(\ref{eq:Mstarnum}, \ref{eq:Fstarnum}). The meson-LECs will be taken from the MILC2010 set, see Tab.~\ref{tab:Li}. The two combinations $\tilde{b}_{D,F}$ will then be fixed to the corresponding values determined in the previous subsection, for each of the four sets of $b_{1-11}$ from Tab.~\ref{tab:bi}. 

With given values for $\tilde{b}_{D,F}$ for $\sqrt{2B_{0}\bar{m}}=0.42\,\mathrm{GeV}$, we can determine seven more parameters from the present data set. First, the $(\delta m_{\ell})^{2}$-terms in the baryon mass formulae can be fixed by four free parameters, which may be taken as
\begin{align}\label{eq:d1tilde}
\tilde{d}_{1} := d_{1}-\frac{1}{3}d_{3}\,,\quad d_{2}\,,\quad \tilde{d}_{4}:= d_{4}+\frac{4}{3}d_{3}
\end{align}
(see e.g. eq.~(\ref{eq:fB2})) and $\tilde{d}_{7}$ defined in eq.~(\ref{eq:deflinpar}). Furthermore, inserting 
\begin{align}
b_{D}=\tilde{b}_{D}-8(0.42\,\mathrm{GeV})^{2}d_{3}
\end{align}
in eq.~(\ref{eq:mstar}), we can determine the combinations
\begin{align}
\begin{split}
b_{0}'&:= b_{0}-\frac{16}{3}(0.42\,\mathrm{GeV})^{2}d_{3}\,,\\
\tilde{d}_{6} &:= \frac{4}{9}d_{3}+d_{6}+\frac{1}{3}d_{7} \label{eq:b0prime} 
\end{split}
\end{align}
and $m_{0}$ from the running of $X_{N}(\delta m_{\ell}=0)$ with $B_{0}\bar{m}$. The combination $\tilde{b}_{0}$ used in the previous subsection can then be computed, for $\sqrt{2B_{0}\bar{m}}=0.42\,\mathrm{GeV}$, as 
\begin{align}
\tilde{b}_{0}=b_{0}'+12B_{0}\bar{m}\tilde{d}_{6}\,.
\end{align}
The results are given below, in Tab.~\ref{tab:res}. The fits are very good, $\chi^{2}/\mathrm{d.o.f.}\sim 0.2$. This can be understood from the fact that the fan plots look very linear near the symmetric point, and the linear approximation fixed by the parameter combinations $\tilde{b}_{D,F}$ already describes this part quite well - better than one could have expected beforehand. Moreover, we have enough free LECs at hand to describe the behavior of $X_{N}$ in a satisfying manner. The emerging fitting parameters are all of natural size, and $m_{0}\sim 1\,\mathrm{GeV}$. In the fits leading to the results in Tab.~\ref{tab:reswexp}, we have also included the physical (experimental) values for the baryon masses in the fit, assigning an error of $5\,\mathrm{MeV}$ to each due to our neglection of isospin-breaking effects. The baryon masses at the physical point are obtained inserting $\nu=-0.885$ and our standard value for $2B_{0}\bar{m}$ in our mass formulae. As a consistency check of our procedure, we have also performed the fits without the constraints on $\tilde{b}_{D,F}$. The corresponding results can be found in Tab.~\ref{tab:resall} and Tab.~\ref{tab:resallwexp}. To get an estimate of the stability of our results due to uncertainties in the input values and higher order corrections, we have also performed full fits where the meson decay constant is strictly truncated to the value $F_{0}$ of the MILC2010 set of LECs, see Tab.~\ref{tab:Li}. The fit parameters following with this input can be read off from Tab.~\ref{tab:resallwexpF0milc}. We observe a good overall agreement of all these fits: varying the input parameters $b_{i},F_{\star},L_{i}$ always leads to similar results. We will discuss our results in detail in the next section.
\begin{table}[h]
\caption{Results for the fit parameters. The set-number refers to Tab.~\ref{tab:bi}. $\tilde{b}_{D,F}$ are fixed input. $m_{0}$ is given in $\mathrm{GeV}$, $b'_{0},\tilde{b}_{D,F}$ in $\mathrm{GeV}^{-1}$, and the $d_{i}$ are given in $\mathrm{GeV}^{-3}$ at a scale $\mu=1150\,\mathrm{MeV}$. Here, the experimental masses are not included in the fit.}
\label{tab:res}
\begin{center}
\begin{ruledtabular}
\begin{tabular}{c c c c c c c c c c c}
 set & $\tilde{b}_{D}$ & $\tilde{b}_{F}$ & $m_{0}$ & $b'_{0}$ & $\tilde{d}_{1}$ & $d_{2}$ &  $\tilde{d}_{4}$ &  $\tilde{d}_{6}$ & $\tilde{d}_{7}$ & $\tilde{b}_{0}$ \\
\hline
 1 & 0.078 & -0.352 & 1.029 & -0.364 & 0.035 & 0.122 & 0.020 & -0.532 & -0.244 & -0.927  \\
\hline
 2 & 0.062 & -0.355 & 1.051 & -0.373 & 0.040 & 0.126 & 0.018 & -0.603 & -0.272 & -1.011  \\
\hline
 3 & 0.083 & -0.358 & 1.064 & -0.400 & 0.030 & 0.121 & 0.030 & -0.641 & -0.289 & -1.078  \\
\hline
 4 & 0.072 & -0.305 & 0.930 & -0.271 & 0.025 & 0.101 & 0.042 & -0.213 & -0.113 & -0.497  \\
\end{tabular}
\end{ruledtabular}
\end{center}
\end{table}
\begin{table}[h]
\caption{Results for the fit parameters. Here, the fit includes the experimental baryon masses.}
\label{tab:reswexp}
\begin{center}
\begin{ruledtabular}
\begin{tabular}{c c c c c c c c c c c}
 set & $\tilde{b}_{D}$ & $\tilde{b}_{F}$ & $m_{0}$ & $b'_{0}$ & $\tilde{d}_{1}$ & $d_{2}$ &  $\tilde{d}_{4}$ &  $\tilde{d}_{6}$ & $\tilde{d}_{7}$ & $\tilde{b}_{0}$ \\
\hline
 1 & 0.078 & -0.352 & 1.026 & -0.370 & 0.035 & 0.046 & 0.044 & -0.525 & -0.257 & -0.926  \\
\hline
 2 & 0.062 & -0.355 & 1.046 & -0.381 & 0.041 & 0.050 & 0.041 & -0.594 & -0.287 & -1.010  \\
\hline
 3 & 0.083 & -0.358 & 1.058 & -0.411 & 0.030 & 0.047 & 0.054 & -0.630 & -0.306 & -1.077  \\
\hline
 4 & 0.072 & -0.305 & 0.936 & -0.262 & 0.024 & 0.017 & 0.063 & -0.222 & -0.115 & -0.496  \\
\end{tabular}
\end{ruledtabular}
\end{center}
\end{table}
\begin{table}[h]
\caption{Consistency check. Here, the combinations $\tilde{b}_{D,F}$ are not fixed, and the experimental masses are not included in the fit. The values for $\tilde{b}_{0}$, $\tilde{b}_{F}$ are computed for each set of fit results.}
\label{tab:resall}
\begin{center}
\begin{ruledtabular}
\begin{tabular}{c c c c c c c c c c c c}
 set & $m_{0}$ & $b'_{0}$ & $\tilde{b}_{D}$ & $b'_{F}$ & $\tilde{d}_{1}$ & $d_{2}$ &  $\tilde{d}_{4}$ &  $\tilde{d}_{6}$ & $\tilde{d}_{7}$ & $\tilde{b}_{0}$ & $\tilde{b}_{F}$ \\
\hline
 1 & 1.028 & -0.372 & 0.087 & -0.405 & 0.043 & 0.119 & 0.039 & -0.531 & -0.244 & -0.933 & -0.321 \\
\hline
 2 & 1.050 & -0.381 & 0.072 & -0.409 & 0.048 & 0.122 & 0.036 & -0.602 & -0.273 & -1.017 & -0.323 \\
\hline
 3 & 1.063 & -0.407 & 0.092 & -0.412 & 0.040 & 0.121 & 0.051 & -0.640 & -0.289 & -1.084 & -0.327 \\
\hline
 4 & 0.929 & -0.278 & 0.079 & -0.346 & 0.033 & 0.095 & 0.059 & -0.211 & -0.114 & -0.501 & -0.280 \\
\end{tabular}
\end{ruledtabular}
\end{center}
\end{table}
\begin{table}[h]
\caption{Consistency check. Here, the fit includes the experimental baryon masses, but the combinations $\tilde{b}_{D,F}$ are not fixed.}
\label{tab:resallwexp}
\begin{center}
\begin{ruledtabular}
\begin{tabular}{c c c c c c c c c c c c}
 set & $m_{0}$ & $b'_{0}$ & $\tilde{b}_{D}$ & $b'_{F}$ & $\tilde{d}_{1}$ & $d_{2}$ &  $\tilde{d}_{4}$ &  $\tilde{d}_{6}$ & $\tilde{d}_{7}$ & $\tilde{b}_{0}$ & $\tilde{b}_{F}$ \\
\hline
 1 & 1.021 & -0.378 & 0.077 & -0.357 & 0.039 & 0.022 & 0.051 & -0.518 & -0.257 & -0.926 & -0.341 \\
\hline
 2 & 1.042 & -0.389 & 0.062 & -0.361 & 0.044 & 0.025 & 0.048 & -0.587 & -0.287 & -1.010 & -0.343 \\
\hline
 3 & 1.054 & -0.417 & 0.082 & -0.365 & 0.035 & 0.025 & 0.062 & -0.624 & -0.305 & -1.077 & -0.347 \\
\hline
 4 & 0.930 & -0.269 & 0.069 & -0.295 & 0.027 & 0.010 & 0.067 & -0.213 & -0.115 & -0.494 & -0.302 \\
\end{tabular}
\end{ruledtabular}
\end{center}
\end{table}
\begin{table}[h]
\caption{Stability check. Here we use the value from MILC2010 for the meson decay constant as input everywhere, instead of $F_{\star}$. The fit includes the experimental baryon masses, but the combinations $\tilde{b}_{D,F}$ are not fixed.}
\label{tab:resallwexpF0milc}
\begin{center}
\begin{ruledtabular}
\begin{tabular}{c c c c c c c c c c c c}
 set & $m_{0}$ & $b'_{0}$ & $\tilde{b}_{D}$ & $b'_{F}$ & $\tilde{d}_{1}$ & $d_{2}$ &  $\tilde{d}_{4}$ &  $\tilde{d}_{6}$ & $\tilde{d}_{7}$ & $\tilde{b}_{0}$ & $\tilde{b}_{F}$ \\
\hline
 1 & 1.116 & -0.329 & 0.118 & -0.470 & 0.075 & 0.121 & 0.134 & -1.247 & -0.524 & -1.649 & -0.385 \\
\hline
 2 & 1.144 & -0.337 & 0.097 & -0.476 & 0.084 & 0.129 & 0.123 & -1.411 & -0.583 & -1.831 & -0.385 \\
\hline
 3 & 1.161 & -0.371 & 0.126 & -0.482 & 0.067 & 0.127 & 0.157 & -1.500 & -0.618 & -1.960 & -0.392 \\
\hline
 4 & 0.993 & -0.211 & 0.098 & -0.381 & 0.048 & 0.063 & 0.152 & -0.516 & -0.250 & -0.757 & -0.336 \\
\end{tabular}
\end{ruledtabular}
\end{center}
\end{table}
\section{Discussion}
\label{sec:Diskussion}
\begin{figure}[h]
\centering
\includegraphics[width=13cm]{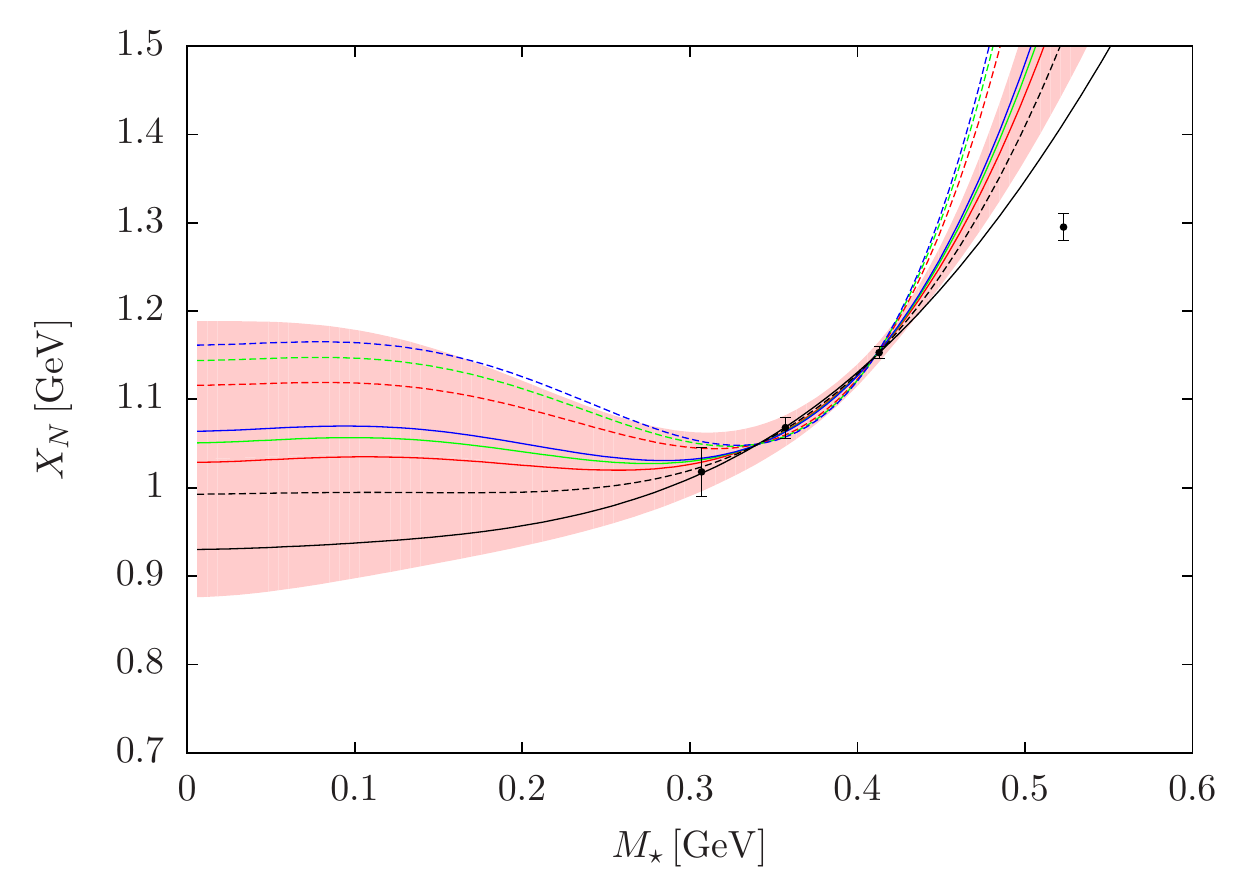} 
\caption{The function $m_{\star}(M_{\star})$ for the four parameter sets from Tab.~\ref{tab:res} (full lines) and from Tab.~\ref{tab:resallwexpF0milc} (dashed lines). Red: set 1, green: set 2, blue: set 3, black: set 4. The band shown here is the one-sigma error band pertaining to the full red curve.}
\label{fig:XNS8sets}
\end{figure}
\subsection*{Convergence at the symmetric point}
We start the discussion of our results with the singlet sector where $\delta m_{\ell}=0$.
Let us have a look at the expansion of $m_{\star}$, using the parameter sets from Tab.~\ref{tab:res} as typical examples. The dimensionless suppression factor could naively be expected to be of the order
\begin{align}
\frac{\sqrt{2B_{0}\bar{m}}}{4\pi F_{\star}}\sim\frac{420}{4\pi\cdot 112}\sim 0.3\,.
\end{align}
Of course, we already know, e.g. from the considerations in sec.~\ref{sec:q3}, that the leading nonanalytic loop correction $\sim\,M^{3}$ is much enhanced w.r.t. this expectation.
Using eq.~(\ref{eq:mstar}), we find
\begin{align*}
m_{\star} &=  m_{0} - 4B_{0}\bar{m}(3b_{0}+2b_{D}) - (2B_{0}\bar{m})^{3/2}\frac{(5D^{2}+9F^{2})}{24\pi F_{\star}^{2}} + \frac{(2B_{0}\bar{m})^{2}}{3(4\pi F_{\star})^{2}}\biggl[\,...\,\biggr] +\mathcal{O}(\bar{m}^{5/2})\\
&\rightarrow (1029 + 330 - 397 + 186)\,\mathrm{MeV} = 1148\,\mathrm{MeV}\qquad\textrm{for set 1 from Tab.~\ref{tab:res}},\\
&\rightarrow (1051 + 351 - 397 + 143)\,\mathrm{MeV} = 1148\,\mathrm{MeV}\qquad\textrm{for set 2 from Tab.~\ref{tab:res}},\\
&\rightarrow (1064 + 365 - 397 + 115)\,\mathrm{MeV} = 1147\,\mathrm{MeV}\qquad\textrm{for set 3 from Tab.~\ref{tab:res}},\\
&\rightarrow \,\,(\, 930 + 236 - 397 + 380)\,\mathrm{MeV} = 1149\,\mathrm{MeV}\qquad\textrm{for set 4 from Tab.~\ref{tab:res}}.
\end{align*}
The result for $m_{\star}$ is very close to  $m_{\star}^{num}$ in all cases due to the constraint included in the fit. The coefficient in square brackets giving the fourth order contribution can be read off from eq.~(\ref{eq:mstar}). We see that the observed convergence behavior is still inconclusive, though the fourth order contributions employing the LECs $b_{1-11}$ from the meson-baryon scattering amplitudes are roughly in line with a suppression factor of $\sim 0.5$. A similar pattern (at physical quark masses) has e.g. been found in \cite{Lehnhart:2004vi}, where some fourth order effects were estimated. In contrast, the fits for $b_{1-11}=0$ result in expansions of $m_{\star}$ which seem meaningless in the sense of perturbation theory. 

To further study the convergence issue, we set the average quark mass to $2B_{0}\bar{m}=(300\,\mathrm{MeV})^{2}$ (this entails $F_{\star}\rightarrow 104\,\mathrm{MeV}$ using eq.~(\ref{eq:Fstar})). Then we would have
\begin{align*}
m_{\star} &\rightarrow (1029 + 169 - 168 - 10)\,\mathrm{MeV} = 1020\,\mathrm{MeV}\qquad\textrm{for set 1 from Tab.~\ref{tab:res}},\\
m_{\star} &\rightarrow (1051 + 179 - 168 - 39)\,\mathrm{MeV} = 1023\,\mathrm{MeV}\qquad\textrm{for set 2 from Tab.~\ref{tab:res}},\\
m_{\star} &\rightarrow (1064 + 186 - 168 - 57)\,\mathrm{MeV} = 1025\,\mathrm{MeV}\qquad\textrm{for set 3 from Tab.~\ref{tab:res}},\\
m_{\star} &\rightarrow ( 930 + 120 - 168 + 122)\,\mathrm{MeV} = 1004\,\mathrm{MeV}\qquad\textrm{for set 4 from Tab.~\ref{tab:res}}.
\end{align*}

The third order term is still comparatively large, cancelling most of the second order corrections, but the fourth order contributions already seem to be well under control for this low average quark mass. Unfortunately, there is no data available for such low average quark masses yet.

The parameters which are mostly determined from the behavior of the singlet sector can therefore not be fixed very accurately from the data set considered here. In particular, $m_{0},b_{0}'$ and $\tilde{d}_{6}$ will be subject to large uncertainties. This also reflects itself in the following experiment: Including also the fourth data point for $m_{\star}(M_{\star})$ at $M_{\star}\sim 530\,\mathrm{MeV}$ in the fit, we observe that the aforesaid three parameters vary rather drastically (e.g. the results of Tab.~\ref{tab:res} shift to $m_{0}\sim 800\,\mathrm{MeV}$, $b_{0}'\sim -0.9\,\mathrm{GeV}^{-1}$, $\tilde{d}_{6}\sim 0.04\,\mathrm{GeV}^{-3}$), while the remaining parameters remain roughly the same. This is particularly disturbing for $m_{0}$, a constant that reappears everywhere in three-flavor chiral extrapolations for baryon observables. We would like to point out that it would be very helpful for this purpose to have some more $m_{\star}$-data points for $M_{\star}<300\,\mathrm{MeV}$. Up to now, we can only give the following very rough bounds, based on Tab.~\ref{tab:res}-\ref{tab:resallwexpF0milc} and the variation of the input parameters:
\begin{align}
 800\,\mathrm{MeV} \leq & \,m_{0}  \leq 1200\,\mathrm{MeV}\,,\label{eq:m0fin}\\
 -1\,\mathrm{GeV}^{-1} \leq & \,b_{0}'  \leq 0\,\mathrm{GeV}^{-1}\,,\label{eq:b0pfin}\\
 -2\,\mathrm{GeV}^{-3} \leq & \,\tilde{d}_{6}  \leq 0.50\,\mathrm{GeV}^{-3}\,.\label{eq:d6tilfin}
\end{align}
As a consequence of the results in Tab.~\ref{tab:resallwexpF0milc}, we revise the lower limit for the allowed range of the combination $\tilde{b}_{0}$ (see eq.~(\ref{eq:b0tillin})),
\begin{align}\label{eq:b0tilfin}
-2\,\mathrm{GeV}^{-1} \leq  \,\tilde{b}_{0}  \leq -0.20\,\mathrm{GeV}^{-1}\,.
\end{align}
In Fig.~\ref{fig:XNS8sets}, we show $m_{\star}$ as a function of $M_{\star}$. One immediately sees that the uncertainty grows rapidly above the region where the curves are fixed by the fit to data (note that the fourth data point shown here is not included in the fits). Recall that {\em the only difference} between the fits leading to the full and the dashed curves in this figure is the $\mathcal{O}(p^{5})$-effect $F_{\star}\rightarrow F_{0}$. Beyond $M_{\star}\gtrsim 500\,\mathrm{MeV}$, this higher-order effect has a very large impact on $m_{\star}$. Of course one could obtain good fits to lattice data in this region with both choices for the decay constant, but in our opinion, these fits would not be reliable in the sense of a stable determination of the fitting parameters (LECs). The finding that chiral extrapolations should not be trusted for meson masses much above $500\,\mathrm{MeV}$ is consistent with related studies \cite{Djukanovic:2006xc,Bernard:2006te}. Beyond this regime, the extrapolations depend very strongly on the details of the input and fine-tuning of input and fitting parameters.
\subsection*{Symmetry breaking}
\begin{figure}[h]
\centering
\includegraphics[width=13cm]{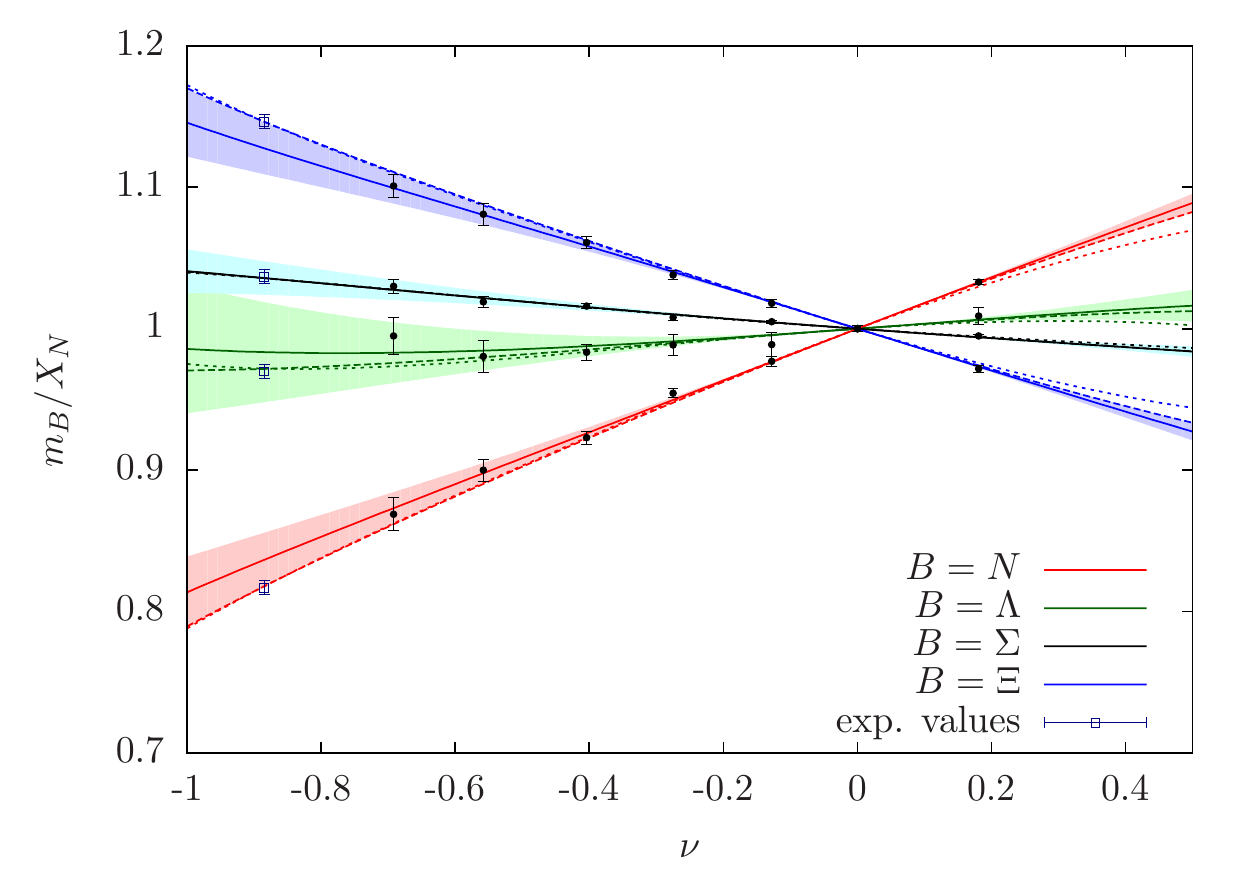} 
\caption{Three different results for the fan plot curves, with $b_{1-11}$ from set 1. Full line: from Tab.~\ref{tab:res}, dashed line: from Tab.~\ref{tab:reswexp}, dotted line: from Tab.~\ref{tab:resallwexp}. The bands shown here are the one-sigma error bands pertaining to the full line. Recall that the fits corresponding to the dashed and dotted lines include the experimental baryon masses.}
\label{fig:fanplot3sets}
\end{figure}
Let us now discuss the parameters which determine the SU(3) symmetry breaking visualized by the fan plots, as seen in Fig.~\ref{fig:fanplot3sets}. First, it is reassuring to observe that it does not matter whether $\tilde{b}_{D,F}$ are used as fixed input in the full fits or not: Tables~\ref{tab:res} and \ref{tab:resall} and Tables~\ref{tab:reswexp} and \ref{tab:resallwexp} are very similar, even though the input values in Tab.~\ref{tab:res} and \ref{tab:reswexp} were obtained with approximations linear in the variable $\nu$ for the ratios $f_{B}$ and a quadratic approximation for $X_{N}$, and truncated chiral expansions for $f_{B}$ (eq.~(\ref{eq:feins})), including a reordering of terms to eliminate $m_{0}$ in favor of $m_{\star}$ in the ratios. The bounds from the linear fits given in eqs.~(\ref{eq:bDtillin},\ref{eq:bFtillin}) are respected in all our resulting parameter sets. Even a rather large variation of the meson decay constant (from $F_{\star}\sim 112\,\mathrm{MeV}$ to $F_{0}\sim 80\,\mathrm{MeV}$) does not invalidate this picture, as can be seen from the fourth and the last column of Tab.~\ref{tab:resallwexpF0milc}. It is also interesting to note that the combination $\frac{20}{3}\tilde{b}_{D}-\tilde{b}_{F}$ falls into the bound given in eq.~(\ref{eq:bDFbound}) for $b_{D,F}$ for all fits from Tabs.~\ref{tab:res}-\ref{tab:resallwexp}, and is slightly above that range for the fits with set 1-3 from Tab.~\ref{tab:resallwexpF0milc}. From all these observations we can conclude that the determination of the LEC combinations parameterizing the linear part of the flavor symmetry breaking is very stable, thanks to the impressive new data leading to the fan plots. This is also illustrated by the error band and the three different fit curves shown in Fig.~\ref{fig:fanplot3sets}.\\ It remains to discuss the parameters determined from the nonlinear symmetry breaking effects. They appear at fourth chiral order in our formulae, which is the highest order we can compute exactly within the present one-loop framework. Therefore, we cannot expect a very high precision here. From our results, we extract the following allowed ranges for the remaining parameter combinations (in units of $\mathrm{GeV}^{-3}$):
\begin{align}
 -0.05 \leq & \,\tilde{d}_{1}  \leq 0.15\,,\label{eq:d1tilfin}\\
  0.00 \leq & \,d_{2}  \leq 0.25\,,\label{eq:d2fin}\\
 -0.10 \leq & \,\tilde{d}_{4}  \leq 0.20\,,\label{eq:d4tilfin}\\
 -0.75 \leq & \,\tilde{d}_{7}  \leq -0.05\,.\label{eq:d7tilfin}
\end{align}
We stress again that these estimates concern the numerical values of the renormalized LECs at $\mu=1150\,\mathrm{MeV}$, and should be evolved by eqs.~(\ref{eq:Renorm}) when using infrared regularization at a different scale $\mu$. 

\begin{figure}[h]
\centering
\includegraphics[width=10cm]{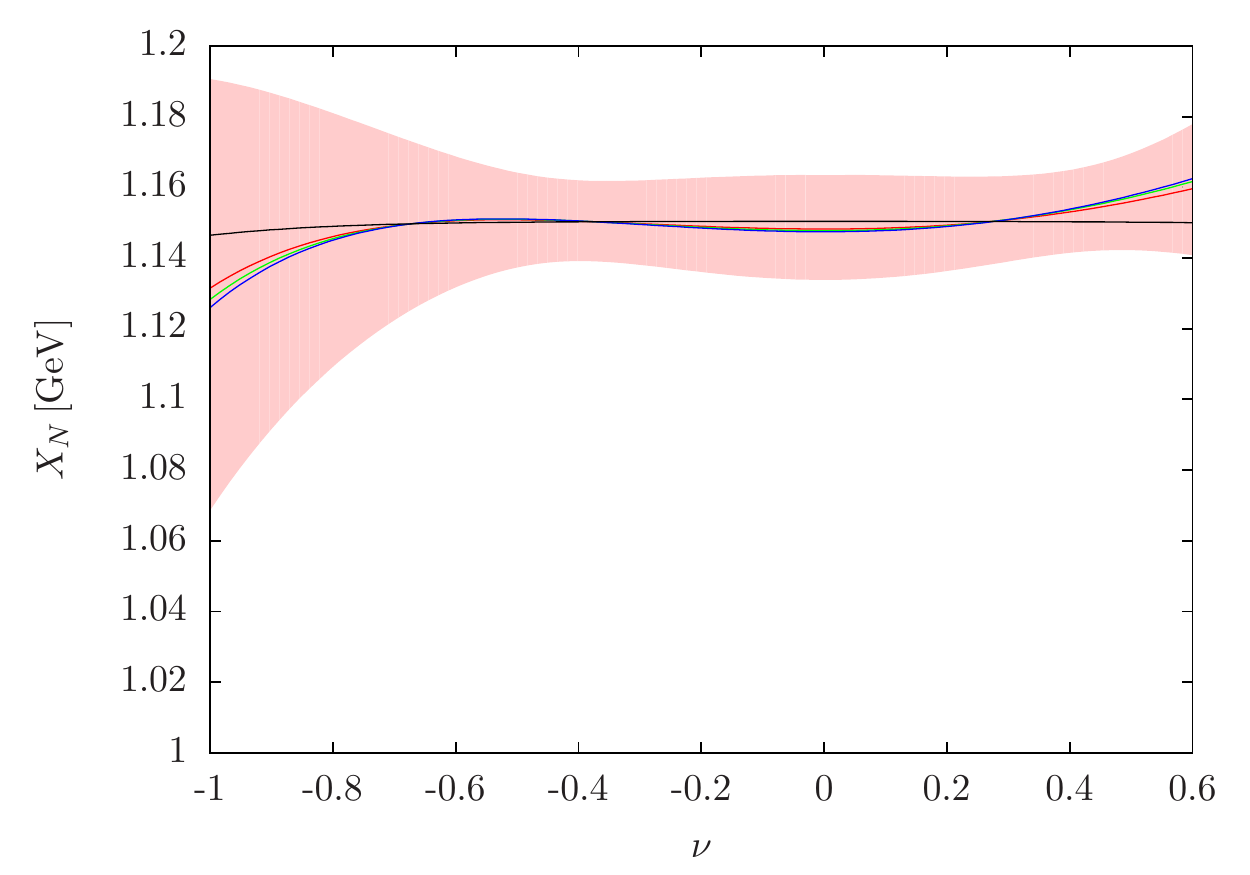} 
\caption{The function $X_{N}(\nu)$ for the four fits of Tab.~\ref{tab:res}. Red: set 1, green: set 2, blue: set 3, black: set 4.  The band shown here is the one-sigma error band pertaining to the red line.}
\label{fig:XNofnu}
\end{figure}

Obviously, the purely linear fits lead to an underestimation of the uncertainty in $\tilde{d}_{7}$: Here we had to shift the lower limit of our previous constraint, mostly due to the fit results shown in Tab.~\ref{tab:resallwexpF0milc}, compare eqs.~(\ref{eq:d7tillin}) and (\ref{eq:d7tilfin}). In Fig.~\ref{fig:XNofnu}, we show the constrained behavior of the function $X_{N}(\nu)$, which is mostly secured by the combination $\tilde{d}_{7}$. Here, the expectation of convergence at low orders {\em must} fail because the higher order terms in eq.~(\ref{eq:XN2}) have to compensate for the leading ($\mathcal{O}(p^{3})$) effect. This might partly explain why the range for $\tilde{d}_{7}$ in eq.~(\ref{eq:d7tilfin}) turns out to be comparatively broad.

Including the experimental masses in the fit mainly (and strongly) affects the result for $d_{2}$, which parameterizes the curvature of the fan plot curves for the $\Xi$ and the nucleon, compare e.g. Tab.~\ref{tab:res} and Tab.~\ref{tab:reswexp}, and the full versus the dashed and dotted curves in Fig.~\ref{fig:fanplot3sets}. While in all cases $d_{2}$ is of natural size, this shows that a reliable chiral extrapolation to the physical point, using one-loop BChPT expressions together with fan plot data, must still be considered problematic, in spite of the improvements in determining many LEC combinations worked out in the present contribution. 
In the framework of Heavy-Baryon ChPT, the low-energy constants $d_{1-7}$ have been estimated from the resonance saturation principle, combined with a fit to experimental masses and sigma terms, by Borasoy and Mei{\ss}ner \cite{Borasoy:1996bx}, see their eq.~(68). Setting $\beta_{R}=0$ in those equations, and forming our preferred combinations, we see that their results for $\tilde{d}_{1,4,7}$ and $d_{2}$ all lie in the ranges given above (their $\tilde{d}_{6}$, however, is much smaller than in our tables, $\tilde{d}_{6}^{BM}=-0.022\,\mathrm{GeV}^{-3}$). Moreover, according to their estimates, $\tilde{b}_{D}^{BM}=0.176\,\mathrm{GeV}^{-1}$ is somewhat above our allowed range (see eq.~(\ref{eq:bDtillin})), while their $\tilde{b}_{F}^{BM}=-0.344\,\mathrm{GeV}^{-1}$ again agrees very well with our typical fit results. One should note that such a comparison can only be of qualitative nature, for various reasons. Besides the differences of the schemes and the issue of scale dependence, it is not known how accurate the estimates from resonance saturation can probably be here, in particular because these LECs are not of ``dynamical'' nature (see also the discussion in sec.~5.2 of \cite{Frink:2004ic}). What is more, the masses and $\sigma_{\pi N}(0)=45\,\mathrm{MeV}$ at the physical point have been input in the estimates of \cite{Borasoy:1996bx}, and we have already seen that in general the convergence properties of the chiral expansions are far from satisfying there due to the large $m_{s}$-corrections.\\
In \cite{Durr:2011mp}, several parameterizations for the baryon masses were used to test the model dependence of the resulting fits to lattice data, and of the determination of the sigma terms. While this model dependence is found to be an important source of systematic uncertainty, we note that the LECs $b_{0,D,F}$ obtained from the $\mathcal{O}(p^{3})$-BChPT approach in \cite{Durr:2011mp} agree very well with the values we find for the corresponding parameters $\tilde{b}_{0,D,F}$ in our analysis.

\subsection*{Note added}
Shortly after a first version of the present work appeared on the web, two other analyses of the lattice data discussed here became available \cite{Lutz:2012mq,Ren:2012aj}. Of those, \cite{Ren:2012aj} uses a framework similar to the one used here (which is essentially given in \cite{Frink:2004ic}). It therefore seems reasonable to attempt a comparison of the resulting fitting parameters, though there is still a minor difference in the renormalization schemes used. Let us first have a look at the most important parameter in the singlet sector, the baryon mass in the chiral limit. The range of possible values for this parameter we found is quite broad, see eq.~(\ref{eq:m0fin}). In Tab.~\ref{tab:resm0} we show their result in the first column, together with those of three other references in which this parameter is determined. We see that only the result of \cite{MartinCamalich:2010fp} is not in line with our broad estimate.
\begin{table}[h]
\caption{Results for $m_{0}$ (in GeV) from ref.~\cite{Ren:2012aj} and three other references. For our estimate, see eq.~(\ref{eq:m0fin}).}
\label{tab:resm0}
\begin{center}
\begin{ruledtabular}
\begin{tabular}{c c c c c}
 Ref. & \cite{Ren:2012aj}, Fit I & \cite{MartinCamalich:2010fp} & \cite{Semke:2011ez} & \cite{Borasoy:1996bx} \\
\hline
 $m_{0}$ & $0.880\pm 0.022$ & $0.756\pm 0.032$ & $0.944\pm 0.002$ & $0.767\pm 0.110$ \\
\end{tabular}
\end{ruledtabular}
\end{center}
\end{table}
Going on to the other LECs, and forming the combinations determined here with the results given in Table 6 (Fit I) of \cite{Ren:2012aj}, we note that they are all consistent with the ranges we gave above, at least within the errors given there, with one exception, which is given by $\tilde{b}_{D}$ (we have checked that the shift from $\mu=1\,\mathrm{GeV}$ used in the latter reference to the value $\mu=1.15\,\mathrm{GeV}$ used here does not induce large effects on our ranges). However, one should note that the parameter $b_{D}$ in \cite{Ren:2012aj} jumps from a small value $\sim 0.05/\mathrm{GeV}$ at $\mathcal{O}(p^{2})$ and  $\mathcal{O}(p^{3})$ (consistent with our bounds on $\tilde{b}_{D}$) to a comparatively large value $\sim 0.22/\mathrm{GeV}$ in the $\mathcal{O}(p^{4})$ fits. We suppose that this parameter is particularly afflicted by convergence problems, given that a large part of the data set analysed in \cite{Ren:2012aj} are at the border, or even outside, of the region where the three-flavor expansions work in a reliable manner.
 
\subsection*{Sigma term}
Rewriting our formula for the contact term contribution to the sigma term, eq.~(\ref{eq:sigPiNcont2}), in terms of our fitting parameters defined in eqs.(\ref{eq:deflinpar},\ref{eq:b0prime}), we find 
\begin{align}
\sigma_{\pi N}^{(ct)}(0) = -2B_{0}m_{\ell}\biggl(4b'_{0}+2\tilde{b}_{D}+2\tilde{b}_{F}+96B_{0}\bar{m}\tilde{d}_{6}+\frac{16}{3}(B_{0}\delta m_{\ell})(3\tilde{d}_{1}+3d_{2}-4d_{3}+9d_{5}+6\tilde{d}_{7})\biggr)\,.
\end{align}
This shows that we can write the sigma term resulting from our formulae as
\begin{align}\label{eq:sigPiNd35}
\begin{split}
\sigma_{\pi N}(0) &= \sigma_{\pi N}(0)|_{d_{3}=d_{5}=0} + \frac{8}{3}(2B_{0}\delta m_{\ell})(2B_{0}m_{\ell})(9d_{5}-4d_{3})\\ 
&\approx \sigma_{\pi N}(0)|_{d_{3}=d_{5}=0} - \frac{8}{3}M_{\pi}^{2}(X_{\pi}^{2}-M_{\pi}^{2})(9d_{5}-4d_{3})\,, 
\end{split}
\end{align}
where the first term is written in terms of our fitting parameters. The unknown constants $d_{3},d_{5}$ parameterize the dependence on $\bar{m}$ of the terms linear in $\delta m_{\ell}$. The latter dependence starts as soon as one includes chiral loops, i.e. at $\mathcal{O}(p^{3})$, see e.g. eq.~(\ref{eq:feins}). Let us consider two examples to estimate the unknown parameter combination in question, ignoring the issue of convergence for the moment. For the fit of Table~\ref{tab:res} with set 1, the first term is $\sigma_{\pi N}(0)|_{d_{3}=d_{5}=0}=67\,\mathrm{MeV}$, while for the fit with set 4, it results in $47\,\mathrm{MeV}$. For the remainder, we can only give a very rough estimate based on a consistency condition: Requiring that the terms $\sim d_{3},d_{5}$ do not yield more than a $100\%$ correction in the combinations $\tilde{b}_{D},b'_{F}$ gives bounds $\left|d_{3}\right|\lesssim 0.07\,\mathrm{GeV}^{-3}$ and $\left|d_{5}\right|\lesssim 0.38\,\mathrm{GeV}^{-3}$ $\Rightarrow -0.41\lesssim(d_{5}-\frac{4}{9}d_{3})\,\mathrm{GeV}^{3}\lesssim 0.41$, which produces sigma terms between $39\ldots 95\,\mathrm{MeV}$ for set 1 and $19\ldots 75\,\mathrm{MeV}$ for set 4. These bounds are consistent with recent three-flavor lattice determinations, see e.g. \cite{Durr:2011mp,Horsley:2011wr}. We also mention that the above consistency bounds for $d_{3,5}$ are obeyed by the results for these two parameters given in \cite{Ren:2012aj}.

However, it is to be noted that we do not see a clear sign of convergence for the sigma term at physical quark masses, so one should not overinterprete these last results. Basically we have to conclude that the sigma term could only be reliably determined from BChPT extrapolations when more data also for smaller average quark masses (so that e.g. $M_{\star}\sim 300\,\mathrm{MeV}$) become available.
\section{Conclusions}
In summary, we have analyzed lattice data for octet baryon masses from \cite{Bietenholz:2011qq} employing manifestly covariant BChPT with three dynamical flavors. We were able to give bounds on the numerical values of certain combinations of low-energy constants (LECs), which are, in our opinion, more reliable and accurate than it was possible before data for fixed average quark mass became available in \cite{Bietenholz:2011qq}: As we have seen, the simulation strategy used in that work offers some advantages for this purpose. Eqs.~(\ref{eq:bDtillin},\ref{eq:bFtillin}), (\ref{eq:m0fin},\ref{eq:b0pfin},\ref{eq:d6tilfin}) and (\ref{eq:d1tilfin},\ref{eq:d2fin},\ref{eq:d4tilfin},\ref{eq:d7tilfin}) should be considered as our main results. We hope that these bounds will be useful in the near future, e.g. when studying the chiral behavior of other hadron observables with the same simulation strategy. Though we have fixed the running of the baryon masses on the two trajectories $\lbrace\bar{m}=\mathrm{const.},\delta m_{\ell}\rbrace$ and $\lbrace\bar{m},\delta m_{\ell}=0\rbrace$, there are still two undetermined parameters, chosen here as $d_{3}$ and $d_{5}$, which would have to be fixed in order to make an $\mathcal{O}(p^{4})$ prediction for the sigma term. However, even if those parameters were fixed accurately, the uncertainty in the determination of the sigma term would still be very large due to the very slow convergence near the physical point. For an investigation of sigma terms, a two-flavor version of ChPT is certainly superior; we refer to \cite{Bali:2012qs} for a recent determination. In closing, we stress again that additional data points for lower average quark masses would be very helpful in order to reach a truly controlled chiral extrapolation.
\acknowledgments{We acknowledge discussions and useful communications with G.~Bali, M.~Frink, M.~Mai, U.-G.~Mei{\ss}ner, P.~E.~L.~Rakow, A.~Rusetsky and A.~Sternbeck. This work was supported by the Deutsche Forschungsgemeinschaft
SFB/Transregio 55. L.~G.~ acknowledges support by the European Union under Grant Agreement number 256594 (IRG).}

\bibliographystyle{apsrev}

\end{document}